\documentclass[11pt]{article}
\usepackage{amsmath, amssymb}
\usepackage{lipsum}
\usepackage{listings}
\usepackage[hidelinks]{hyperref}
\usepackage{fontspec}
\usepackage{graphicx} 
\usepackage{pgffor}
\usepackage{authblk}
\usepackage{subcaption} 
\usepackage[rightcaption]{sidecap}
\usepackage{caption}
\usepackage{hyperref}
\usepackage{bm}
\usepackage[table]{xcolor} 
\usepackage{algorithm}
\usepackage{algpseudocode}
\usepackage{helvet}

\usepackage{booktabs}
\usepackage{comment}
\usepackage{soul}
\usepackage[backend=biber,style=numeric]{biblatex}
\usepackage[margin=1in]{geometry}

\usepackage{fancyhdr}
\title{\vspace{-1cm}\textbf{Estimating Complex Densities using Two-Stage Normalizing Flows}}
\author[1]{Roxana Darvishi}
\author[1]{David C. Stenning}
\author[2]{Ted von Hippel}
\author[1]{Owen G. Ward}

\affil[1]{Department of Statistics and Actuarial Science, Simon 
Fraser University}
\affil[2]{Department of Physical Sciences and SARA,  Embry-Riddle Aeronautical University}

\affil[ ]{roxana\_darvishi@sfu.ca, david\_stenning@sfu.ca, vonhippt@erau.edu, owen\_ward@sfu.ca}
\date{\today} % No date

\usepackage{abstract}

\definecolor{light}{HTML}{FDF0D5}
\definecolor{lighter}{HTML}{FEF8EC}
\definecolor{dblue}{HTML}{003049}
\definecolor{lblue}{HTML}{669BBC}
\definecolor{dred}{HTML}{780000}
\definecolor{dorange}{HTML}{D57443}
\definecolor{lightGray}{HTML}{B5B5B5}
\definecolor{note}{HTML}{a7c957}
\definecolor{appendix}{RGB}{223, 234, 247}

\newcommand{\teta}{\bm{\theta}}

\newcommand{\y}{\bm{y}}

\newcommand{\R}{\mathbb{R}}
\newcommand{\E}{\mathbb{E}}

\newcommand{\bu}{\bm{u}}
\newcommand{\bU}{\bm{U}}
\newcommand{\bX}{\bm{X}}
\newcommand{\bx}{\bm{x}}
\newcommand{\by}{\bm{y}}
\newcommand{\fu}{f_U}
\newcommand{\fx}{f_{\bm{X}}}
\newcommand{\Ti}{{T^{-1}}}

\newcommand{\owen}[1]{\textcolor{red}{[Owen: #1]}}

\newcommand{\new}[1]{{\color{lblue}#1}}

\begin{document}
	
	% Apply first page style with no header for logo
	\thispagestyle{plain}
	
	% Title
	\maketitle %\vspace{-2cm}
	
	\begin{abstract}
        In many scientific applications, the target probability distribution cannot be 
        evaluated in closed form or sampled from directly. Instead, it can often be decomposed 
        into multiple components, some of which are accessible only through samples generated 
        by simulators or external datasets, while others admit tractable mathematical 
        expressions or are specified through statistical assumptions about variable 
        relationships. Developing inference methods that coherently integrate these 
        heterogeneous sources of information remains an open challenge.
        In this paper, we propose a Two-Stage Normalizing Flows framework for approximating 
        and sampling from such distributions. The method first learns the densities of 
        components for which only samples are available, and then combines the outputs with 
        the analytically specified terms to reconstruct the full target distribution in a 
        second stage. The resulting model enables both point-wise density evaluation and 
        efficient generation of representative samples, without requiring direct access to 
        the full target density or joint samples from the complete model. 
        We assess the proposed approach through simulation studies in joint density inference 
        and Bayesian hierarchical models with inaccessible likelihoods. 
        The proposed framework is able to accurately recover complex, highly nonlinear target 
        structures using only partial information about the target density, providing
        stable and flexible approximations in settings where standard modeling assumptions do 
        not hold (or when complete access to the target distribution is not available). 
        Analysis of a large scale astronomy application highlights interesting differences
        between our method and existing approaches.
        Our normalizing flows procedure offers a robust and flexible approach to inference for intractable target distributions across both simulated and real-world applications.
        % We also compare the approach with the Two-Stage Fully Bayesian method (should I reference this here?) in a real-data astronomy application, highlighting differences in the smoothness, stability, and structure of the resulting posterior estimates, particularly in settings where the available samples may not fully align with the target distribution.

	\end{abstract}
	
	\section{Introduction}
    Characterizing a probability distribution that describes a system of variables or phenomena under study is a primary objective for many applied statistics problems (e.g. \cite{casella2002statistical}).  Achieving this relies on the ability to either evaluate a density or to generate samples that reflect its underlying morphology, which in turn supports a wide range of inferential tasks including estimation, prediction, simulation, and uncertainty quantification. 
    This distribution has a central role in statistical modeling and inference, serving as the object through which knowledge about the quantities of interest is summarized. 
	We use the term target density to denote the probability distribution associated with the quantities of interest, which serves as the central object we seek to evaluate, approximate, or sample from.
    
	Most existing approaches to probabilistic inference rely on strong assumptions about how the target distribution can be accessed, typically requiring either a set of representative samples or the ability to evaluate its density up to a normalizing constant. However, these conditions are not always met, particularly when the target is complex, partially observed, difficult to sample from, or defined implicitly through intractable functions. To bridge this gap, researchers commonly modify the original target by assuming a tractable family of distributions over unknown components of the target density, or by imposing multiple layers of approximations. Although computationally convenient, such adjustments may oversimplify the problem or lead to inference about a distribution that differs from the true target through the process of making the procedure more computationally convenient.
	
	In many cases, the target density can be expressed as a product of multiple components. For example, in Bayesian hierarchical models, it factorizes into a likelihood term and several prior distributions, whereas in joint density inference it decomposes into the product of marginals and conditionals. In practice, not all these terms may be known explicitly. Instead, only piecewise information about each component may be accessible, in the form of either samples, structural assumptions, or mathematical expressions that outline the relationship between variables. This motivates the need for a flexible framework capable of leveraging the available information in its original form without requiring explicit knowledge of the full target density or joint samples from the complete model.
	
	For instance, this situation arises in settings where the likelihood itself is inaccessible, but a black box software can generate Monte Carlo draws from a distribution that implicitly contains this likelihood term. The objective remains to perform posterior inference by utilizing the likelihood information carried by the software-generated samples.
	Similarly, when the target is a multivariate random vector, it may not be feasible to obtain samples from the full joint distribution, while a set of subvector samples may exist from a separate research study. Although these subvectors may share common components, their realizations are not matched across sources. The goal is to combine the information provided by each subvector sample to recover the full joint structure.
	In other settings, the distribution of interest may include terms that cannot be sampled directly, yet their functional form is known due to theoretical or scientific principles. This includes known marginals, conditionals, and structural assumptions and relationships imposed by domain-specific knowledge. Inference in such cases requires a method that is able to incorporate each analytic components to characterize the density of interest.
	
	These examples illustrate that information about the objective can appear in different representations and from several independent sources, each describing only part of the overall structure. This highlights the need for new statistical methodology that is capable of combining all partial and structurally different knowledge of the target to reconstruct it as a whole, without relying on restrictive assumptions. Such an approach is particularly valuable when high-quality draws from the full target vector cannot be collected, or its density cannot be efficiently computed.
	
	While combining heterogeneous pieces of information is the primary challenge, it is also useful to consider the different ways in which a distribution may be used once it has been characterized.
	Inference with a probability distribution involves two fundamental tasks. One is the ability to draw representative samples. This enables uncertainty quantification, approximation of expectations and integrals, investigation of dependence structure, and simulation-based prediction. The second task is to evaluate the probability density at specific points, which is necessary for likelihood-based parameter estimation, hypothesis testing, and model comparison. Being able to carry out both tasks offers greater flexibility, supporting a wider range of inferential goals within a unified approach without switching frameworks or imposing additional restrictions.
	
	There are two main limitations with standard approaches in this context. First, they assume that the full target distribution is available in a single unified representation, which prevents them from incorporating sample-based information, analytic components, structural assumptions, and distinct datasets that observe only subsets of variables. Second, they are generally designed to support only one of the main inferential tasks. Methods that generate samples require access to the density, whereas the approaches that compute probabilities depend on complete data and explicit likelihoods. Simulations from the entire model are necessary for the procedures that bypass density evaluation, and non-parametric methods do not extend well in higher dimensions. Together, these limitations make most existing methods unsuitable in such settings. 
	
	In this paper, we propose an approach based on Normalizing Flows \cite{papamakarios2021normalizing} \cite{rezende2015variational} for modeling target distributions in the settings described above. Normalizing Flows represent probability distributions through a sequence of invertible transformations applied to a simple base distribution. This class of models is capable of establishing connections between samples and analytic densities, allowing both sampling and density evaluation since the transformations can be applied in either direction.
	
	The proposed method proceeds in two stages. In the first stage, the densities of the sample-based components are approximated using a forward-KL Normalizing Flows model. Subsequently in the second stage, the estimated components are combined with the remaining analytically known terms, recreating the full target distribution in a tractable form. A reverse-KL Normalizing Flows model is then trained on this representation to obtain the overall target distribution. The model in Stage 2 is able to evaluate the target density at any given point as well as generating samples with efficiency comparable to sampling from the base distribution, which is often considered to be a standard normal distribution.
	
	The remainder of the paper is organized as follows. In Section~\ref{sec:background}, we provide background on Normalizing Flows and briefly outline related popular approaches in the literature. Section~\ref{sec:methods} introduces the proposed framework, along with its implementation details and generalizations. In Section~\ref{sec:simulations}, we present two simulation studies demonstrating the ability of the proposed method to capture complex, nonlinear target structures and to outperform existing approaches, using a joint density inference problem and a hierarchical model example. Section~\ref{sec:realdata} demonstrates the practical utility of the approach through an application to an astronomy dataset. Finally, in Section~\ref{sec:conclusion} we summarize the strengths and limitations of the proposed method and discusses future directions.

	\section{Background}
	\label{sec:background}
	
	We first current approaches to probabilistic inference and discuss their limitations in settings where the target distribution is only partially accessible. We then introduce hierarchical models as a motivating class of problems and describe an existing algorithm, called the Two-Stage Fully Bayesian method, developed for hierarchical settings with inaccessible likelihoods, which we later use as a baseline for comparison. We conclude with a brief overview of Normalizing Flows and their training mechanisms, which form the foundation of our proposed approach. 
	
	Standard approaches to inference for complex probability models assume full access to the target distribution, either through direct evaluation of its density or by simulating complete draws of all components of the target vector. In Markov chain Monte Carlo (MCMC) methods \cite{MCMC}, for instance, the transition rules that define the chain require evaluating the unnormalized target density to determine how the chain moves through the parameter space. MCMC samplers produce a sequence of dependent draws with an asymptotic (i.e., stationary) distribution equal to that of the target distribution. Therefore, such samplers are typically run for many iterations such that later draws follow a distribution that is a good approximation of the target distribution. However, when each evaluation of the the target density is computationally expensive, running the MCMC sampler for a sufficient number of iterations may be impractical.
    % \old{ In astrostatistics, MCMC samplers are employed infer cosmological parameters from observational data and to perform Bayesian analysis of large-scale cosmological models} \cite{astro_mcmc_1} \cite{astro_mcmc_2}.

    Variational inference \cite{variationalinference} provides another general framework for Bayesian inference, in which the target distribution is approximated by optimizing over a pre-specified family of distributions. Like MCMC, this approach assumes that the full target density can be evaluated, since the optimization of the evidence lower bound requires computing or approximating expectations involving the target distribution.
	When the structure of the full target distribution is unknown or only partially observed, specifying an appropriate variational family in advance may not be feasible. Moreover, optimizing the evidence lower bound still requires the full density to be accessible, which prevents the method from being applied when some components of the target are known only through samples and others only through analytic expressions.
	
	Nonparametric density estimation techniques \cite{wasserman2006all} such as kernel estimators \cite{silverman1986density} or nearest neighbour based approaches \cite{NN} avoid parametric assumptions but rely entirely on samples. Their major drawback is that their accuracy declines sharply as the dimension of the target vector grows \cite{wasserman2006all}, and they cannot easily incorporate known structural relationships among variables. Moreover, they cannot be used to generate new samples and are often highly sensitive to tuning choices.
	
	Approximate Bayesian computation (ABC) \cite{ABC} avoids likelihood evaluation but depends on generating synthetic data from the complete model and comparing it to the observed data. This requires a model that can produce joint samples of all variables involved in the target vector, making it unsuitable when the observed information consists of separate samples on different subsets of the variables. Furthermore, ABC generally demands a large number of simulations, which is not practical when sampling is computationally expensive.
	
	Taken together, the approaches described above rely on the target distribution being fully available either through direct density evaluation or the ability to simulate draws from full target vector. Moreover, they do not simultaneously provide access to sufficiently representative samples from the target distribution and values of its density at arbitrary points in the target space, which limits the types of inferential questions they can answer. 
    
	 Obtaining samples on the entire target vector may require costly data collection or long simulation runs, while assessing density values may involve high-dimensional integrals or computationally intensive model evaluations. 
    In practice, we have little control over what is known about a distribution of interest. In many applications, knowledge of the target distribution may consist of samples from some of its components together with analytic expressions for others. For example, a simulator may generate draws from some components of the target density, while the other components remain mathematically tractable. In other cases, data from several independent research studies may provide samples from different subvectors of the target without matched joint observations.
	Situations of this kind arise frequently in joint density inference \cite{ESL} \cite{ joint_density_1} \cite{ joint_density_2} \cite{ joint_density_3},
    Bayesian hierarchical models \cite{BDA}, meta-analytic data integration
    \cite{ meta_analysis} \cite{meta_analysis_2}, and other experimental contexts. Addressing such scientific problems requires an approach that can synthesize these sample-based and analytic components without altering the underlying target or imposing restrictive modeling assumptions.
	
	\subsection{Motivating Example: Hierarchical Models}\label{sec:hier_mod}
	Hierarchical models (e.g. \cite{BDA}) are used when the structure of a problem involves multiple layers, where quantities at one level depend on or inform those at another. In these models, parameters themselves are treated as random variables with structured dependence across groups or levels, making them ideal for complex data scenarios. By pooling information across groups, hierarchical models can reduce variance and improve the quality of estimates \cite{Gelman_Hill_2006}. The framework also allows prior knowledge to be incorporated through priors and hyperpriors defined at different levels of the hierarchy. Hierarchical modeling is widely employed in statistics in fields such as biology \cite{hier_biology}, medicine \cite{hier_med}, environmental science \cite{hier_environment}, education \cite{hier_education}, and among other fields, where data are naturally organized into related groups.
	
	As an example, a typical two-level hierarchical model has the following form:
	\begin{align}
		\hspace{0.15\linewidth}
		\begin{split}
			y_{ij} \sim& P_H(y_{ij} \mid \theta_j, \xi_j) \quad i = 1,\dots,n; \; j = 1,\dots,m, \\
			\xi_j \sim& P_H(\xi_j \mid \theta_j), \quad j = 1,\dots,m, \\
			\theta_j \sim& P_H(\theta_j \mid \bm{\psi}), \quad j = 1,\dots,m, \\
			\bm{\psi} \sim& P_H(\bm{\psi}),
		\end{split} \label{hierarchical_model}
	\end{align}
	where $y_{ij}\in\mathbb{R}$ denotes the $i^{\text{th}}$ observation in group $j$, $\theta_j$ and $\xi_j$ are parameters specific to group $j$, and $\bm{\psi}$ is a global parameter vector shared across groups\footnote{Throughout this text we use boldface to denote vectors.}. The term $P_H(y_{ij} \mid \theta_j, \xi_j)$ represents the likelihood, while $P_H(\xi_j \mid \theta_j)$, $P_H(\theta_j \mid \bm{\psi})$, and $P_H(\bm{\psi})$ specify the priors on $\xi_j$ and $\theta_j$, and the hyperprior on $\bm{\psi}$, respectively. The subscript $H$ indicates that these distributions belong to the hierarchical model.
	
	The target distribution for the Bayesian hierarchical model specified in Equation (\ref{hierarchical_model}) is the posterior distribution of all unknown parameters, conditioning on the observed data. By Bayes' theorem, this posterior is proportional to the product of the likelihood and the prior terms, as shown in Equation (\ref{hierarchical_posterior}):
	\begin{equation}
		P(\bm{\theta}, \bm{\xi}, \bm{\psi} \mid \by) \propto
		\left[\prod_{i=1}^{n}\prod_{j=1}^{m} P_H(y_{ij} \mid \theta_j, \xi_j)  \right]
		\left[\prod_{j=1}^{m}P_H(\xi_j \mid \theta_j)P_H(\theta_j \mid \bm{\psi})\right]
		P_H(\bm{\psi}),
		\label{hierarchical_posterior}
	\end{equation}
    where $\bm{\theta} = (\theta_1,\dots,\theta_j)$, $\bm{\xi} = (\xi_1,\dots,\xi_j)$, and $\bm{y}=(y_{11},\dots,y_{1m},y_{21},\dots,y_{2m},\dots,y_{n1},\dots,y_{nm})$.
	
	The posterior representation in Equation (\ref{hierarchical_posterior}) shows that the target distribution is expressed as a product of several components. In many applications, these components are not all accessible in the same format, as some are available only through samples and others are known through tractable mathematical expressions.
	For example, samples from the likelihood may be accessible, but the conditional relationships, such as $P_H(\xi_j \mid \theta_j)$ and $P_H(\theta_j \mid \bm{\psi})$, may be derived from established scientific theory. Most existing methods cannot combine information in mixed formats, such as a combination of samples and analytic terms, within a single hierarchical model.

	In this paper, we focus on hierarchical models with inaccessible likelihoods, referring to situations in which the posterior distribution is difficult to evaluate directly. This may occur when a closed-form expression of the posterior distribution is not available, when its values are generated by a black-box mechanism that prevents evaluation at arbitrary points, or when the density is computationally intractable or costly to evaluate. Specifically, we consider the case where the likelihood is either unknown or intractable, but a "black box" software can generate Monte Carlo samples from 
	\begin{equation}
		P_S( \bm{\theta}, \bm{\xi} \mid \by) \propto P_H(\by \mid \bm{\theta}, \bm{\xi}) P_S(\bm{\theta}, \bm{\xi}),
		\label{software_likelihood}
	\end{equation}
	which represents the original likelihood combined with a software-specific prior $P_S(\bm{\theta}, \bm{\xi}) = P_S(\teta)P_H(\bm{\xi}\mid\teta)$. 
	We assume that the remaining terms in the hierarchical model are known explicitly.
	
	\subsection{Two-Stage Fully Bayesian Algorithm}
	The Two-Stage Fully Bayesian (TSFB) algorithm, first
    proposed by \textcite{thesis}, is an existing approach designed for this particular setup described in Section~\ref{sec:hier_mod}. The method provides a way to generate samples from the posterior in 
    Equation~(\ref{hierarchical_posterior}) without requiring direct evaluation of the original likelihood, but it requires that a sample from the posterior distribution presented in Equation~(\ref{software_likelihood}) is available.
	
	The TSFB algorithm begins by drawing a new value of $\bm{\psi}$ from its conditional distribution $P_H(\bm{\psi} \mid \teta)$ under the hierarchical model in Equation~(\ref{hierarchical_model}). Then, for each group $i$, a proposal for $(\theta_i, \xi_i)$ is selected from the software-generated sample and accepted or rejected using a Metropolis-Hastings step. The likelihood term cancels out in the calculation of the acceptance ratio, allowing the update to be performed without the need to directly evaluate the likelihood function.
    If proper convergence can be achieved, given the constraints imposed by using the software-generated draws to form the proposal distribution in a Metropolis-Hastings scheme, then the result is an approximate sample from the full posterior distribution in Equation~(\ref{hierarchical_posterior}).
    The detailed procedure is provided in Section~\ref{sec:TSFB_alg} of the appendix.
	
	Although effective in this specific context, the TSFB method cannot extend to more general cases for which the posterior does not admit  the structure assumed above, where software-generated samples are available from a distribution proportional to the likelihood combined with a software-specific prior, and the remaining model components are analytically specified. Furthermore, convergence of the Metropolis-Hastings step requires a relatively large number of proposals, imposing a substantial computational burden when generating software draws is time-consuming. Lastly, the TSFB algorithm can be strongly influenced by the software-specific prior and the samples this generates.
    That is, if draws of $(\teta,\bm{\xi})$ from the sampler are far from the likelihood and the posterior distribution under the hierarchical model, it is likely that the resulting samples are not representative of the target distribution. This sensitivity to the choice of software prior is investigated in Section \ref{sec:hier-example}.
	
	\subsection{Normalizing Flows}
	
	Normalizing Flows \cite{papamakarios2021normalizing} \cite{rezende2015variational} are a family of generative models that provide a general mechanism for defining expressive probability distributions. 
	Starting from a simple base distribution, such as a standard normal, a sequence of transformations is applied to obtain an arbitrarily complex target density. Detailed overviews on this framework are available in \cite{papamakarios2021normalizing} and \cite{kobyzev2021normalizing}. In what follows, we provide a brief introduction to Normalizing Flows, focusing on the aspects most relevant to the proposed methodology.
	%The central idea is to begin with a simple base distribution, such as a standard normal distribution, and transform it into a more complex target density through a sequence of transformations.
	
	Adopting the notation in \cite{papamakarios2021normalizing}, let $\bU$ and $\bX$ be continuous $\R^D$-valued random variables representing the base and target distributions, respectively. Then, consider
	\begin{equation}
		\bX = T(\bU), \; \text{where} \;  \bU \sim \fu(\bu),
		\label{NF_1}
	\end{equation}
	where transformation $T$, parameterized by the vector $\bm{\phi}$, is required to be invertible, with both $T$ and its inverse $T^{-1}$ being differentiable. This allows us to express the target density via the change of variables formula:
	\begin{equation}
		\fx(\bx) = \fu(\Ti(\bx)) \left|\det \left(\frac{\partial \Ti}{\partial \bx}\right) \right|.
		\label{NF_2}
	\end{equation}
	By the chain rule for derivatives and the properties of Jacobians, the transformation $T$ can be expressed as a composition of multiple mappings $T = T_1 \circ \dots \circ T_k$, allowing the construction of arbitrarily complex densities through successive compositions of tractable transformations. In practice, $T$ and $\Ti$ are modeled using neural networks if the Jacobian determinant can be computed efficiently.
	%as long as the Jacobian determinant is efficient to compute.
	A Normalizing Flows model enables two main operations: sampling via Equation~(\ref{NF_1}) by mapping the base distribution samples to the target space, and density evaluation through Equation~(\ref{NF_2}) by calculating the inverse transformation and its Jacobian determinant.
	
	A flow-based model denoted by $ f^*(\bx; \bm{\phi})$ aims to optimize for parameters $\bm{\phi}$ for which the divergence or discrepancy between $\fx(\bx)$ and $ f^*(\bx; \bm{\phi})$ is minimized. A common choice for the loss function is the
	Kullback-Leibler (KL) divergence \cite{KL}. Depending on the available information, a forward or reverse KL Normalizing Flows can be implemented. Each approach uses a different representation of the loss function to incorporate the available information (i.e. target samples or exact density values).
	%which will be discussed in the following sections.
	A properly trained Normalizing Flows model cangenerate new samples and perform point-wise density evaluation simultaneously, regardless of whether a forward or reverse KL model was fit \cite{papamakarios2021normalizing}.
	
	\subsubsection{Forward KL Normalizing Flows}
	The forward KL divergence is well-suited for situations in which we have the ability to generate samples from the target distribution, but we cannot necessarily evaluate its density. For a constant $C$ that does not depend on $\bm{\phi}$, the forward KL divergence in this setup can be written as:
	\begin{align}
		\mathcal{L}(\bm{\phi}) 
		&= D_{KL}\left[ \fx(\bx) \lVert f^*(\bx; \bm{\phi}) \rVert \right] \nonumber \\ 
		&= -\E_{\fx(\bx)}\left[\log f^*(\bx;\bm{\phi})\right] + C  \label{forwardKL}\\ 
		&= -\E_{\fx(\bx)}\left[\log \fu(T_{\bm{\phi}}^{-1}(\bx)) 
		+ \log \left| \det J_{T_{\bm{\phi}}^{-1}}(\bx)\right|\right]+ C. \nonumber
	\end{align}
	
	Assuming we have a set of samples $\{\bx_n\}_{n=1}^N$ from $\fx(\bx)$, after pushing each sample through the inverse flow, we can form a Monte Carlo estimate of the expectation over $\fx(\bx)$, written as
	\begin{align}
		\mathcal{L}(\bm{\phi})
		&\approx
		-\frac{1}{N}\sum_{n=1}^N
		\left[
		\log \fu(T_{\bm{\phi}}^{-1}(\bx_n))
		+ \log \left| \det J_{T_{\bm{\phi}}^{-1}}(\bx_n) \right|
		\right]
		+ C
	\end{align}
	where the first term is the likelihood of the sample under the base measure and the second term, sometimes called the log-determinant or volume correction, accounts for the change of volume induced by the transformations of Normalizing Flows \cite{kobyzev2021normalizing} 
	\cite{forward_KL_Application}.

	\subsubsection{Reverse KL Normalizing Flows}
	If the true target density can be computed at any point, a Normalizing Flows model can be fitted by minimizing the reverse KL divergence:
	\begin{align}
		\mathcal{L}(\bm{\phi})
		&= D_{KL}\!\left[ f^*(\bx;\bm{\phi}) \,\middle\|\, \fx(\bx) \right] \nonumber \\
		&= \E_{f^*(\bx; \bm{\phi})}\!\left[\log f^*(\bx;\bm{\phi}) - \log \fx(\bx)\right] \label{reverseKL} \\
		&= \E_{\fu(\bu)}\!\left[\log\fu(\bu) - \log\left| \det J_{T_{\bm{\phi}}}(\bu)\right| - \log \fx(T_{\bm{\phi}}(\bu))\right]. \nonumber
	\end{align}
	
	Equation~(\ref{reverseKL}) can be used even if we cannot sample from $\fx(\bx)$. It can also be applied when the true target is known up to a multiplicative constant $C$, since in that case $\log C$ will be an additive constant in Equation~(\ref{reverseKL}).
	
	The reverse KL divergence is often used for variational inference \cite{wainwright2008graphical} \cite{variationalinference} and model distillation \cite{ParallelWaveNet} in which a Normalizing Flows model is trained to replace a target distribution whose density can be evaluated but which may be computationally expensive or difficult to sample from.
	
	\begin{figure}[ht!]
		\centering
		\includegraphics[width=\textwidth]{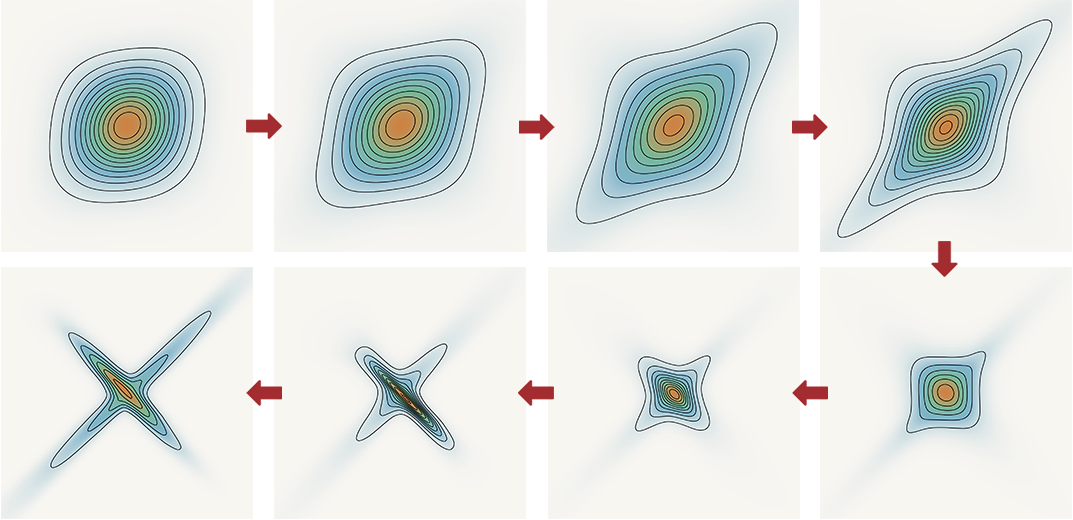}
		\caption{
			Illustration of the training process of a Normalizing Flows model. 
			Starting from a simple base distribution (typically a standard normal), 
			a sequence of invertible transformations gradually morphs the density into the complex shape of the target distribution. 
		}
		\label{fig:nf_learning_sequence}
	\end{figure}
	
    % ----------------------------------------------
    \section{Methodology}\label{sec:methods}
	As discussed in the preceding sections, complete information about the distribution of interest may not always be available. The target density may be unknown, intractable, unavailable in closed form, or require evaluations that involve high-dimensional integrals or other computationally intensive numerical operations.
	Moreover, direct sampling from the target distribution may not be feasible because generating the draws can be difficult, and simulation procedures may themselves produce samples from an approximation of the target or require substantial runtime.
	Nevertheless, the goal is to generate samples from the target distribution, evaluate its density at specific points, or to achieve both tasks for subsequent inference.
	
	In such settings, piecewise information might be available from different sources that can be combined to reconstruct the target density. The proposed method addresses this challenge by providing a framework using flow-based models to integrate all the partial knowledge of the target distribution in the format they are available, either as samples or mathematical expressions, to infer the target density.
	The proposed Two-Stage Normalizing Flows framework,
    described below, is suitable for constructing tractable models for both density evaluation and sample generation for inference under these conditions.
	
	\subsection{Two-Stage Normalizing Flows Framework}
	
	We consider a $D$-dimensional real-valued random variable $X$ with realization 
    $\bx = (x_1, \dots, x_D)^\top \in \mathbb{R}^D$. 
	For each $i = 0,\dots,m$, let $I_i \subseteq \{1,\dots,D\}$ be an ordered index set with 
    $d_i = |I_i|$ such that $\bigcup_{i=0}^{m} \{j: j\in I_i\} = \{1, \dots, D\}$, meaning that these index sets together cover all coordinates.
    For $i = 0,\dots, m$, we define subvectors of $\bx$ by:
	$$ \by_i := (x_{j_1}, \dots, x_{j_{d_i}}),
	\; I_i =  \left( j_1 , \dots, j_{d_i}\right). $$
	The order in the indexing sets $I_i$ does not necessarily correspond to any inherent ordering of the elements in $\bx$. We assume that the target density $\fx(\bx)$ can be factorized as:
	\begin{equation}
		\fx(\bx) \propto h(\by_0)g_1(\by_1) \dots g_m(\by_m).
		\label{target_decomposition}
	\end{equation}
	Here, each $g_i(\by_i)$ denotes the density of the subvector $\by_i$, corresponding to the components of the target distribution from which samples are available, even if the density itself is unknown, difficult to evaluate, or intractable. In this setup, a sample of size $n_i$ from each $g_i(\by_i)$ is available, potentially from 
    separate datasets or simulation procedures.
	The function $h$ refers to the portion of the target that can be expressed as tractable mathematical expressions, but it does not necessarily need to be a density function.
	
	We begin by fitting $m$ separate forward KL Normalizing Flows models on the available samples from each subvector by minimizing Equation~(\ref{forwardKL}) to obtain 
    approximations of the density components $g_i(\by_i)$, $i = 1, \dots, m$. This procedure forms the first stage of the Two-Stage Normalizing Flows framework. The resulting fitted densities, denoted by 
    $\hat{g}_i(\by_i)$, can be evaluated at any point in the target space.
    In the second stage, these estimated components are combined with the analytically specified term $h(\by_0)$ to form the approximated target density
	\begin{equation}
		\tilde{f}_X(\bx) \propto h(\by_0)\prod_{i=1}^{m}\hat{g}_i(\by_i).
		\label{approximated_target}
	\end{equation}
	Since $\tilde{f}_X$ can be computed proportional to a constant at any point in $\R^D$, a reverse KL Normalizing Flows model can be trained to learn the full target distribution by minimizing Equation~(\ref{reverseKL}). An overview of the Two-Stage Normalizing Flows framework is provided in Algorithm~(\ref{alg:two_stage_flow}).

	\begin{algorithm}[ht!]
		\caption{Two-Stage Normalizing Flows Framework}
		\label{alg:two_stage_flow}
		\begin{algorithmic}[1]
			
			\State \textbf{Inputs:}
			\State For each $i=1,\dots,m$, a set of $n_i$ samples $\{\by_i^{(1)},\dots,\by_i^{(n_i)}\}$ from the subvector $\by_i \in \mathbb{R}^{d_i}$.
			\State A tractable function $h(\by_0)$.
			
			\Statex
			\State \textbf{Stage 1: Fit sample-based components via forward KL}
			\For{$i = 1$ \textbf{to} $m$}
			\State \textbf{Inputs:} $\{\by_i^{(1)},\dots,\by_i^{(n_i)}\}$.
			\State Let $g_i(\by_i)$ denote the unknown density of $\by_i$ samples.
			\State Specify an invertible flow transformation $T_{i,\bm{\phi}}:\mathbb{R}^{d_i}\to\mathbb{R}^{d_i}$ with inverse $T_{i,\bm{\phi}_i}^{-1}$ and base density $f_{U_i}$, inducing a flow density $\tilde g_i(\,\cdot\,;\bm{\phi}_i)$ on $\mathbb{R}^{d_i}$.
			\State \parbox[t]{0.87\linewidth}{Train by minimizing the forward KL objective in Equation (\ref{forwardKL}):}
			\[
			\bm{\phi}_i^\star \in \arg\min_{\bm{\phi}_i}\; D_{KL}\!\left(g_i \,\|\, \tilde g_i(\,\cdot\,;\bm{\phi}_i)\right).
			\]
			\State \parbox[t]{0.87\linewidth}{Output: $\hat g_i(\by_i)=\tilde g_i(\by_i;\bm{\phi}_i^\star)$, which can be evaluated point-wise on its support.}
			\EndFor
			
			\Statex
			\State \textbf{Stage 2: Fit the full target via reverse KL}
			\State \textbf{Inputs:} $h(\by_0)$ and $\{\hat g_i(\by_i)\}_{i=1}^m$.
			\State Construct the (unnormalized) approximated target:
			\[
			\tilde{f}_X(\bx) \propto h(\by_0)\prod_{i=1}^{m}\hat g_i(\by_i).
			\]
			\State Specify an invertible flow transformation $T_{\bm{\phi}} : \mathbb{R}^D \to \mathbb{R}^D$ with inverse $T_{\bm{\phi}}^{-1}$ and base density $f_U$. Define $q(\cdot;\phi)$ as the pushforward of $f_U$ under $T_{\bm{\phi}}$.
			\State Train by minimizing the reverse KL objective in Equation (\ref{reverseKL}):
			\[
			\phi^\star \in \arg\min_{\bm{\phi}}\; D_{KL}\!\left(q(.;\bm{\phi}) \,\|\, \tilde{f}_X\right).
			\]
			\State \parbox[t]{0.87\linewidth}{Output: $\hat f(\bx) = q(\bx;\bm{\phi}^*)$}
			\Statex
			\State \textbf{Use Cases:}
			\begin{itemize}
				\item \textbf{Generate samples:} draw $u \sim f_U$, set $\bx = T_{\bm{\phi}^\star}(u)$.
				\item \textbf{Evaluate the density at $\bx$:} compute $\hat f(\bx) $
			\end{itemize}
			
		\end{algorithmic}
	\end{algorithm}

	To generate samples from the target distribution, we draw $\bu$ from the base distribution $\fu$ and map it through the learned transformation $T_{\phi^*}$, which produces a sample in the target space. This implies that sampling from the target can be carried out almost as efficiently as generating a sample from the base distribution, which is typically considered to be a standard Normal distribution.
	
	Density evaluation at an arbitrary point $\bx^\prime$ in the target space can be performed in two ways. The first approach uses the factorized representation in Equation~(\ref{approximated_target}), using the fitted $\hat{g}_i$ densities from the output of the forward KL Normalizing Flows models in Stage 1. The second uses the model from Stage 2, which provides $\tilde f_X(\bx')$ directly through the change-of-variables formula in Equation (\ref{NF_2}) associated with $T^{-1}_{\phi^*}$. In both approaches, obtaining the target density does not involve expensive calculations or intractable terms. This capability facilitates estimation of integrals and expected values, and enables assessment of distributional structure, uncertainty quantification, and hypothesis testing.
	
	\subsection{Implementation}
	
	The main objective of the forward KL models is to approximate the unknown density 
    components that will later be combined together to approximate the overall target 
    density. For this reason, it is crucial to ensure that the transformation layers used 
    in the 
    Normalizing Flows models at this stage have stable and well behaved Jacobian 
    determinants \cite{ jacobian_stability}, \cite{ jacobian_stability2}.
    In particular, each Jacobian determinant needs to be well defined throughout the 
    target space,
	remain finite, and should not change drastically within a small neighborhood, 
    ensuring that the resulting $\hat{g}_i$ density estimates are numerically stable and 
    vary smoothly rather than displaying abrupt local fluctuations.
	
	This property is essential because the estimated densities are used directly in constructing Equation~(\ref{approximated_target}), which forms the basis for training the reverse KL flow in Stage 2. Training in this stage involves adjusting the parameters $\bm{\phi}$ of the final transformation $T_{\bm{\phi}}$ while evaluating this expression across many regions of the target space. During training, it is common to encounter 
    lower density regions of the $\hat{g}_i$ components,
    and if any of these estimates from Stage 1 yields non-finite or extreme values, the reverse KL objective in 
    Equation~(\ref{reverseKL}) can be undefined, causing optimization to fail.
    Even in high density regions, if the estimated components are unbounded, numerically 
    unstable or fluctuate sharply in small neighborhoods, the gradients passed to the 
    Stage 2 model will be noisy which negatively impacts the optimization dynamic, 
    resulting in unstable updates and poor convergence. Therefore, it is important that 
    when designing the transformation architecture in Stage 1 flow models, stability is 
    prioritized over flexibility by utilizing fewer transformations consisting of simple 
    layers, which are guaranteed to provide well behaved Jacobian determinants. 
	
	Fitting Normalizing Flows model in this framework can be implemented through PyTorch \cite{PyTorch} using existing Normalizing Flow toolkits, including nflows \cite{nflows}, normflows \cite{normflows}, and the flow modules available in Pyro \cite{pyro} and TensorFlow Probability \cite{tfp}. The training of both stages proceeds using stochastic gradient optimization \cite{StochasticApproximation} \cite{MLOptimization} \cite{adam}, and the implementation can be adapted to either CPU or GPU environments.

	\subsection{Normalizing Flows for Hierarchical Models with inaccessible likelihoods}\label{sec:NF_hierarchical}
	We now apply this framework to the motivating example introduced in Section~\ref{sec:hier_mod} concerning hierarchical models with inaccessible likelihoods. For the hierarchical model specified in Equation (\ref{hierarchical_model}), software generated MCMC draws are available from the density in Equation (\ref{software_likelihood}). In this setup, the software prior\new(s?) $P_S(\teta, \bm{\xi})$, $P_H(\teta\mid \bm{\psi})$, and $P_H(\bm{\psi})$ are known analytically, while the remaining terms are either intractable or unavailable. By setting $\by_0 = (\teta, \bm{\xi}, \bm{\psi})$ and $\by_1 = (\teta, \bm{\xi})$, the method proceeds as follows:
	\begin{enumerate}
		\item \textbf{Stage 1:} Train a forward KL Normalizing Flows model to learn the density
		$$g(\by_1) \propto P_H(\bm{Y} \mid \teta, \bm{\xi}) P_H(\bm{\xi}\mid \teta)P_S(\teta)$$
		corresponding to the software samples. Denote the fitted density by $\hat{g}(\teta,\bm{\xi})$.
		\item \textbf{Stage 2:} With the target density being the posterior distribution as specified below,
		$$P(\bm{\theta}, \bm{\xi}, \bm{\psi} \mid \by) \propto
		P_H(\bm{Y} \mid \teta, \bm{\xi}) 
		P_H(\bm{\xi}\mid \teta)P_H(\teta \mid \bm{\psi})
		P_H(\bm{\psi})$$
		let
		$$h(\by_0) = \frac{P(\bm{\theta}, \bm{\xi}, \bm{\psi} \mid \by)}{g(\teta,\bm{\xi})}=
		\frac{P_H(\teta \mid \bm{\psi})P_H(\bm{\psi})}{P_S(\teta)}$$
		Train a reverse KL Normalizing Flows model for the target density
		$$\tilde{f}(\bm{\theta}, \bm{\xi}, \bm{\psi}) \propto \hat{g}(\teta,\bm{\xi})h(\teta, \bm{\xi}, \bm{\psi})$$
	\end{enumerate}
	If the hierarchical structure contains multiple groups, Stage 1 can be repeated for each group to obtain the sampler densities. The product of the fitted densities can be replaced with the single $ \hat{g}$ in the definition of $\tilde{f}$.
	
	\subsection{Generalization}
	The discussion so far has focused on cases where the target density appears as the product of sample-based and analytical terms as in Equation (\ref{target_decomposition}). However, the method can be extended to more general settings. Suppose we have $m$ sample–based components $g_1, \dots, g_m$, where for each $g_i$ we observe $n_i$ samples of the associated subvector $\by_i$, and the remaining analytically available information about the target is summarized into a function $h(\by_0)$. The method can be applied if there exists a known general function $F$ such that the target density $f_X(\bx)$ can be expressed as
	\begin{equation}
		\fx(\bx) \propto F(h(\by_0), g_1(\by_1), \dots, g_m(\by_m))
		\label{general_form_of_target_decomposition}
	\end{equation}
	Then, the Two-Stage Normalizing Flows framework can be implemented by replacing Equation (\ref{general_form_of_target_decomposition}) as the objective distribution in Stage 2 without modifying the procedure in Stage 1.
	
	Examples so far considered $F$ to be a product function. However, it could be replaced with a sum operator or a complex deterministic composition. For example, in the hierarchical structure in Equations (\ref{hierarchical_model}), the unknown likelihood may be expressed as a mixture distribution with known weights
	$$P_H(\bm{Y} \mid \teta, \bm{\xi}) = \pi_1 g_1(\teta, \bm{Y}) + \pi_2g_2(\bm{\xi}) + \pi_3g_3(\teta,\bm{\xi})$$
	where samples are available from the individual mixture components. In such cases, all sample-based components can be learnt through the first stage of the algorithm and recombined through $F$ to approximate the full target density in Stage 2.
	
    % ----------------------------------------------
	\section{Simulation Studies}\label{sec:simulations}
	In this section, we present two simulation studies reflecting the partial-information scenarios that arise in practice and showing how the Two-Stage Normalizing Flows framework reconstructs the target distribution under these constraints. The first examines a joint density inference problem, demonstrating that the proposed approach can efficiently learn complex, highly non-linear target densities. The second considers a hierarchical model with inaccessible likelihood, allowing a direct comparison with the Two-Stage Fully Bayesian algorithm introduced in Section \ref{sec:hier_mod}. The results 
    of these simulations
    indicate that our method consistently outperforms the Two Stage Fully Bayesian algorithm in terms of recovering the correct posterior shape and range, capturing the bivariate structure, and achieving higher accuracy in parameter estimation while operating on a comparable computational timescale. All simulation studies in this section were performed in PyTorch using an Apple M2 Pro CPU. \footnote{Code for the simulations in Section \ref{sec:simulations} and the case study in Section \ref{sec:realdata} is available at \url{https://github.com/Roxanadarvishi/TwoStageNormalizingFlows}.}

	\subsection{Joint Density Inference}
	We first consider a joint density inference problem with a complex, highly non-linear target structure, where $f(x,y,z)$ denotes the joint density of the three-dimensional random vector of interest. To replicate the partial information scenario, we assume that $f(x)$ is analytically known, while information about the remaining components is obtained from two separate simulation procedures, producing samples of $(X,Y)$ and $(X,Z)$
    respectively, potentially of different sizes. 
    This setup mirrors the case in which two separate studies provide observations on different subvectors of the full model. We further assume that $Y$ and $Z$ are conditionally independent given $X$, which is an additional piece of structural information about 
    the underlying structure of this data.
    This example illustrates how our proposed method can incorporate analytic
    expressions, 
    samples generated from different components of the underlying density,
    and structural assumptions to reconstruct a highly curved target density, without the requirement to evaluate the $f(x,y,z)$ directly. 
	
	For this simulation, we specify a known marginal distribution for $X$ with density
	$$
	f(x) = \frac{2}{\Gamma(1/4)}\exp(-x^4).
	$$
	To generate a sample of size $n_1$ from the joint distribution of $(X,Y)$, we draw $X \sim f(x)$, and then simulate $Y\mid X \sim \mathcal{N}(\sin(2X)^3, \sigma^2)$ with $\sigma = 0.1$. Similarly, an independent sample of size $n_2$ from $(X,Z)$ is produced with $Z|X \sim N(\omega\sin(\pi X), \tau^2)$ using parameters $\tau = 0.8$ and 
	$\omega = 5$.
	The target density is the joint distribution of the three variables and is given by
	$$
	f(x,y,z) = f(x)f(y\mid x)f(z \mid x,y) =
	f(x)f(y\mid x)f(z\mid x) = \frac{f(x,y)f(x,z)}{f(x)},
	$$
	according to the conditional independence assumption.
	In the first stage of the proposed framework, we train two forward–KL Normalizing Flow models corresponding to $g_1(x,y) = f(x,y)$ and $g_2(x,z) = f(x,z)$. During training, samples used in each mini-batch are resampled with replacement from the $n_1$ and $n_2$ available pairs. Both models use the same flow architecture, consisting of two layers of Autoregressive Rational Quadratic Spline transformations \cite{durkan2019neuralsplineflows} and 
	ActNorm layers \cite{kingma2018glowgenerativeflowinvertible}.
	These models were trained for 8000 iterations with a batch size of 200 using an Adam optimizer, requiring a run time of 27-28 seconds.
	
	In the second stage, a reverse KL Normalizing Flows model is trained with the following target distribution:
	$$\tilde{f}(x,y,z) = \hat{g}_1(x,y)\hat{g}_2(x,z)h(x), \; \text{ with }\; 
    h(x) = \frac{1}{f(x)}.$$
	The Normalizing Flow model in Stage 2 has the same architecture as before with an added 
    permutation layer \cite{durkan2019neuralsplineflows} and small tuning modifications, running 
    for 25000 iterations in less than 10 minutes.
	
\begin{figure}[ht!]
		\centering
		
		% ---- Marginals ----
		\begin{subfigure}{0.95\textwidth}
			\centering
			\includegraphics[width=\textwidth]{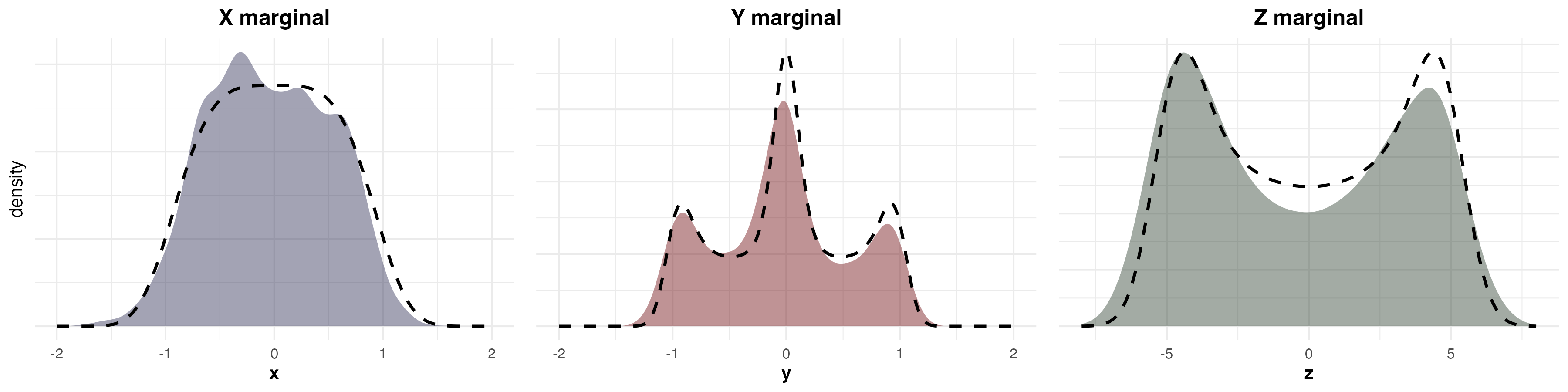}
			\caption{Marginal densities. Shaded regions show the KDE estimates of Stage 2 Normalizing Flow samples , and dashed black curves represent the true marginal densities.}
			\label{fig:marginals_stage2}
		\end{subfigure}
		
		\vspace{0.3cm}
		
		% ---- Bivariates ----
		\begin{subfigure}{0.95\textwidth}
			\centering
            \includegraphics[width=\textwidth]{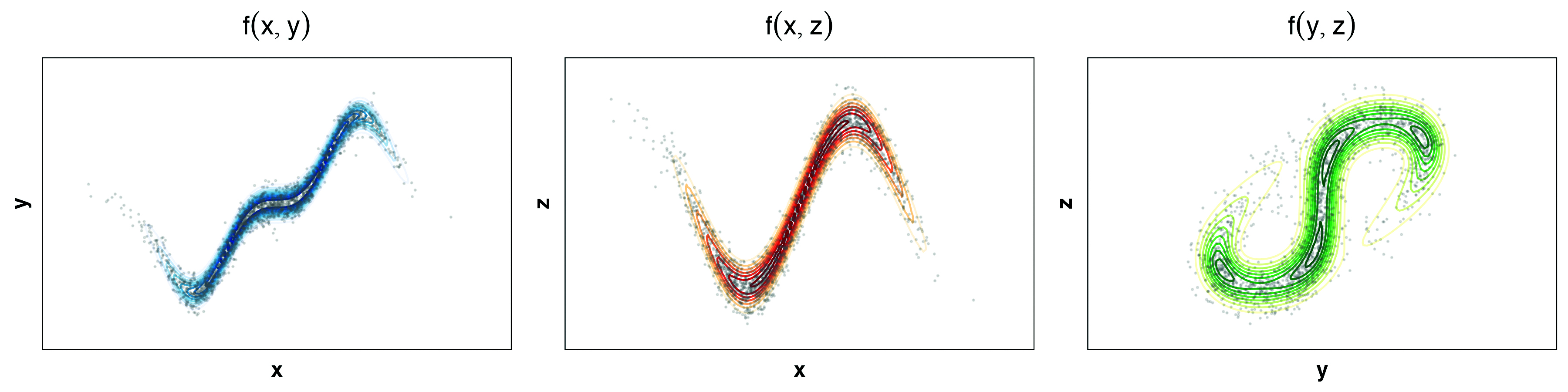}
			\caption{Bivariate densities. Normalizing flow samples are compared against the true $(x,y)$, $(x,z)$, and $(y,z)$ contour levels derived from the true $f(x,y,z)$ model.}
			\label{fig:bivariates_stage2}
		\end{subfigure}
		
		\caption{Comparison of the Stage 2 Normalizing Flow joint model with the ground-truth distribution. The top panel compares marginal fits, and the bottom panel compares bivariate fits.}
		\label{fig:stage2_joint_fit}
	\end{figure}

	Figure \ref{fig:stage2_joint_fit} displays the output of the Stage 2 model alongside the true 
    target marginal density.
	In Figure~\ref{fig:marginals_stage2},
    a Kernel Density Estimate of the samples obtained from the Two Stage Normalizing 
    Flows model aligns well with the true marginals of $X$, $Y$ and $Z$. 

	Figure~\ref{fig:bivariates_stage2} compares bivariate contours, illustrating that the model in Stage 2 captures the complex, non-linear dependence structure across all variable pairs. 
    Despite relying on only a limited set of recycled samples from subvectors of the entire
    target vector, the Stage 2 model provides an accurate fit to the target density within a 
    practical CPU runtime, while supporting both density evaluation and sample generation tasks 
    simultaneously.
	
	\subsection{Simulation Study for Hierarchical Models with Inaccessible Likelihoods}\label{sec:hier-example}
	We next consider a simulation study which mirrors the challenges presented
    by the motivating example of a 
	hierarchical models with intractable likelihood
    discussed in Section~\ref{sec:hier_mod}.
	We assume that a set of software-generated Monte Carlo samples are drawn from a density proportional to the unknown likelihood multiplied by a known software prior.  
    We wish to investigate how the proposed framework performs relative
    to the Two-Stage Fully Bayesian method in Algorithm~(\ref{alg:two_stage_fb}), 
    and examine the influence of the software prior on the resulting inference. 
	When the distribution of software samples is shifted away from the high-density regions of the likelihood, thereby reducing the extent to which these samples reflect the underlying likelihood structure, we show that the Two-Stage Fully Bayesian method fails to capture the target density effectively, while our proposed approach remains robust to the choice of software prior. 
	
	% For the hierarchical model described below
    We consider the following hierarchical model,

    \begin{equation}
		\begin{split}
			&
			Y_i \mid \theta_i \sim N(\theta_i, \sigma^2) \\
			&
			\theta_i \mid \gamma \sim N(\gamma, \tau^2) \\
			&
            \gamma \sim 1
		\end{split} \label{Simulation_Hierarchical_Model_1}
	\end{equation}
	where $i = 1,\dots, J$ for $J$ groups. We assume that the likelihood of the hierarchical model is unknown or inaccessible, while the conditional density of $P_H(\theta_i\mid\gamma)$ and the prior on $\gamma$ are available in closed form. 
    The target density is given by
	\begin{equation}
		P_H(\teta, \gamma \mid \bm{Y}) \propto P_H(\gamma)P_H(\teta \mid \gamma) P_H(\bm{Y} \mid \teta).
		\label{Simulation_Posterior_1}
	\end{equation}
    
	Even though the sampler's density is assumed to be unknown, in the simulation setup we draw the software samples from
	\begin{equation*}
		P_S(\teta\mid \bm{Y}) \propto \prod_{i=1}^J P_H(Y_i\mid \theta_i) P_S(\theta_i),
		\quad \text{ where } \quad
		P_S(\theta_i) \sim N(0,A^2).
    \end{equation*}
    The marginal posterior densities are then given by
    \begin{equation}
		P_S(\theta_i \mid Y_i) \sim N\left(\frac{A^2y_i}{A^2+\sigma^2}, \frac{A^2\sigma^2}{A^2+\sigma^2}\right).
		\label{simulation_software_density_1}
	\end{equation}
	
	Within the Two-Stage Normalizing Flows framework we set 
    $g_1(\teta) = P_S(\teta \mid \bm{Y})$.
	In Stage 1, we train a forward KL flow on software draws, 
    learning a stable estimator $\hat{g}_1(\teta)$ of the unknown density of the software sampler.
    In Stage 2, the learned component $\hat{g}_1(\teta)$ is substituted into Equation (\ref{approximated_target}), with
	$$h(\teta, \gamma) = \frac{P_H(\teta, \gamma \mid \bm{Y}) }{g_1(\teta)} 
	=\frac{P_H(\gamma)P_H(\boldsymbol{\theta}\mid \gamma)}{P_S(\teta)}
	= \prod_{i=1}^J \frac{1}{\sqrt{2\pi}\tau}\exp\left(-\frac{A^2 - \tau^2}{2(A^2 \tau^2)}\left(\theta_i - \frac{A^2\gamma}{A^2-\tau^2}\right)^2\right)$$
	so that the resulting expression is proportional to the posterior density in Equation (\ref{Simulation_Posterior_1}). A reverse KL flow model is then trained on this expression which yields a tractable approximation to the posterior, supporting both density evaluation and efficient sampling.
	
	The Two-Stage Fully Bayesian algorithm is an alternative
    popular approach for inference in this setting.
    The algorithm is first initialized by randomly selecting $\teta^{[TS,0]}$ values with replacement from the pool of software samples. At each iteration $t$, a new value $\gamma^{[TS,t]}$ is drawn from the hierarchical conditional distribution
	$$
	P_H(\gamma\mid \teta^{[TS,t-1]}) \propto P_H(\gamma)P_H(\teta^{[TS,t-1]}\mid \gamma),
	$$
	which reduces to a normal distribution with mean $\bar{\theta}$ and variance $\frac{\tau^2}{J}$. Conditional on $\gamma^{[TS,t]}$, each $\theta_i$ is then updated via a Metropolis–Hastings step, with $P_S(\theta_i\mid Y_i)$ as the proposal and $P_H(\theta_i\mid \gamma^{[TS,t]}, Y_i)$ as the target.
	
    The simulation uses parameter values $J = 3$, $\sigma = 1$, $\tau = 2$, $\gamma = -5$, and software prior scale $A = 0.5$. Stage 1 of the proposed framework approximates the 
    three-dimensional $P_S(\teta\mid \bm{Y})$ using a two-layer flow structure consisting of 
    two Masked Affine coupling blocks and ActNorm transformations. Convergence was achieved after 
    2000 iterations, requiring 3.9 seconds of CPU compute time.
	Kolmogorov–Smirnov tests \cite{KS} comparing marginal samples from Stage 1 model with the software generated draws indicate strong agreement between $\hat{g}_1(\teta)$ and $P_S(\teta \mid \bm{Y})$.
	
	In Stage 2, with the posterior in Equation~(\ref{Simulation_Posterior_1}) as the target 
    density, a reverse KL flow is trained for 5000 iterations, running for roughly 17 seconds.
    The flow is built from three Masked Affine coupling layers and a single ActNorm layer. Under 
    the same simulation conditions,  the Two-Stage Fully Bayesian algorithm completed 30000 
    iterations in 8 seconds. The fitted model can evaluate the posterior at any point within the 
    target space and is able to generate draws from the full posterior almost as efficiently as 
    sampling from a standard normal distribution.
	
	%Goodness-of-fit was assessed by Kolmogorov–Smirnov tests comparing marginal samples from Stage 1 model with the software generated draws. This gave $p$-values of $0.999$, $0.978$, and $0.842$ for $\theta_1$, $\theta_2$, and $\theta_3$ respectively, confirming that $\hat{g}_1(\teta)$ closely aligns with the marginal behavior of $P_S(\teta \mid \bm{Y})$. 
	
	\begin{figure}[h!]
		\centering
        \includegraphics[width=0.99\textwidth]{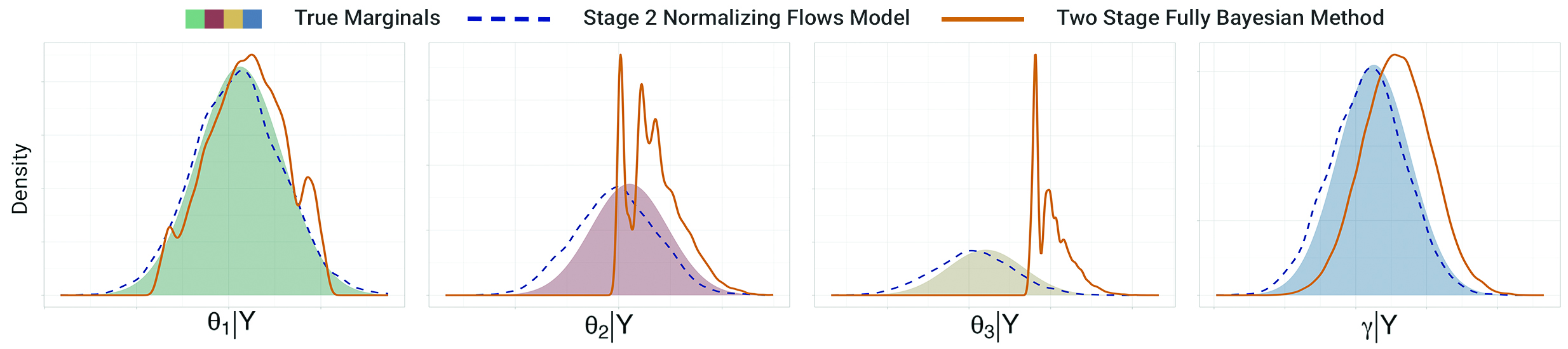}
		\caption{Marginal posterior estimates for each parameter under the proposed Normalizing Flows method and the Two-Stage Fully Bayesian procedure, compared with the true analytic marginals.}
		\label{fig:hierarchical_marginals}
	\end{figure}
	
\begin{figure}[H]
		\centering
		\begin{subfigure}{0.37\textwidth}
			\centering
			\includegraphics[width=\linewidth]{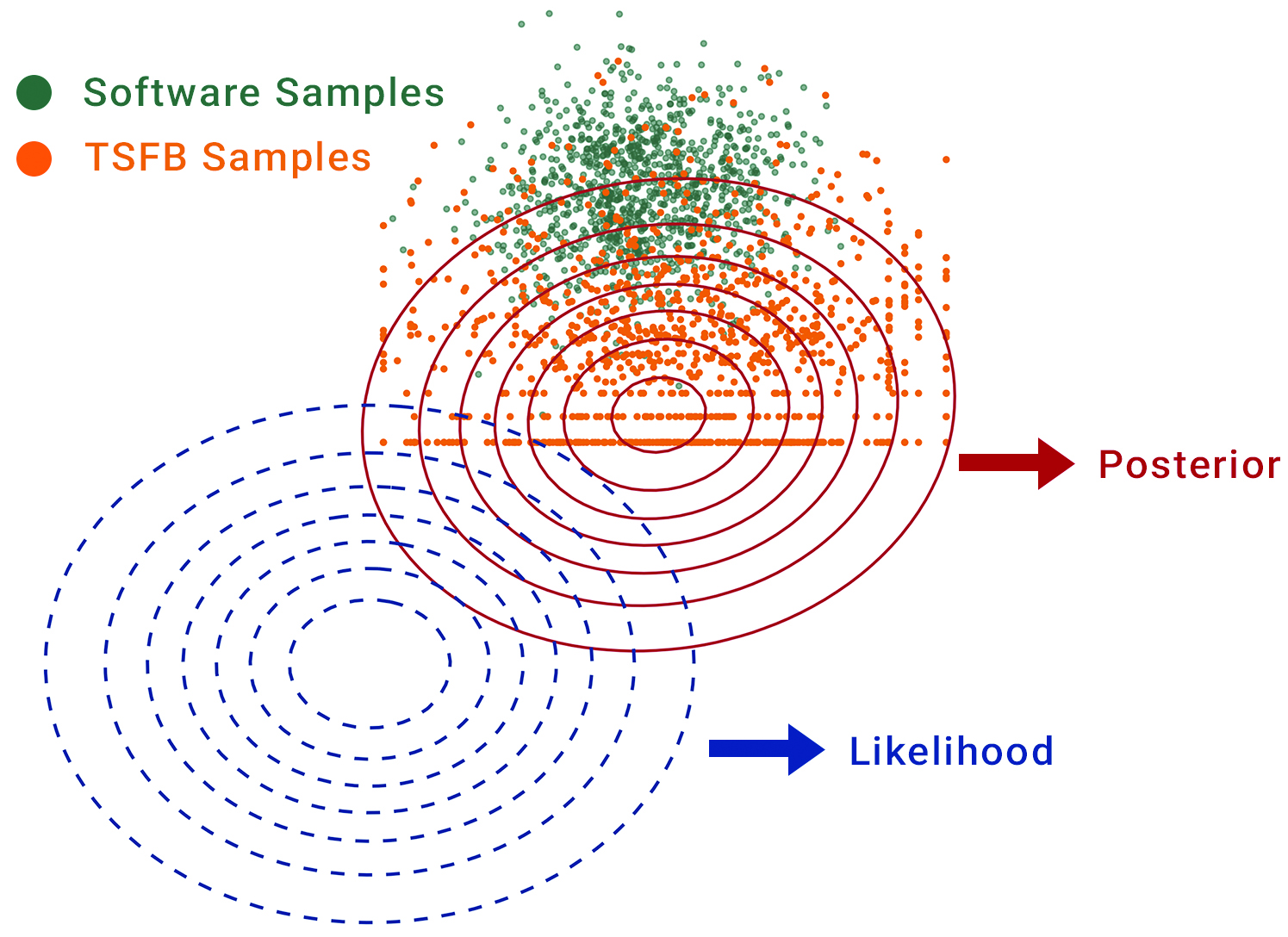}
			\caption{\centering Effect of the sampling prior on software draws.}
            \label{fig:post_comp_a}
		\end{subfigure}
		\hfill
		\begin{subfigure}{0.29\textwidth}
			\centering
			\includegraphics[width=\linewidth]{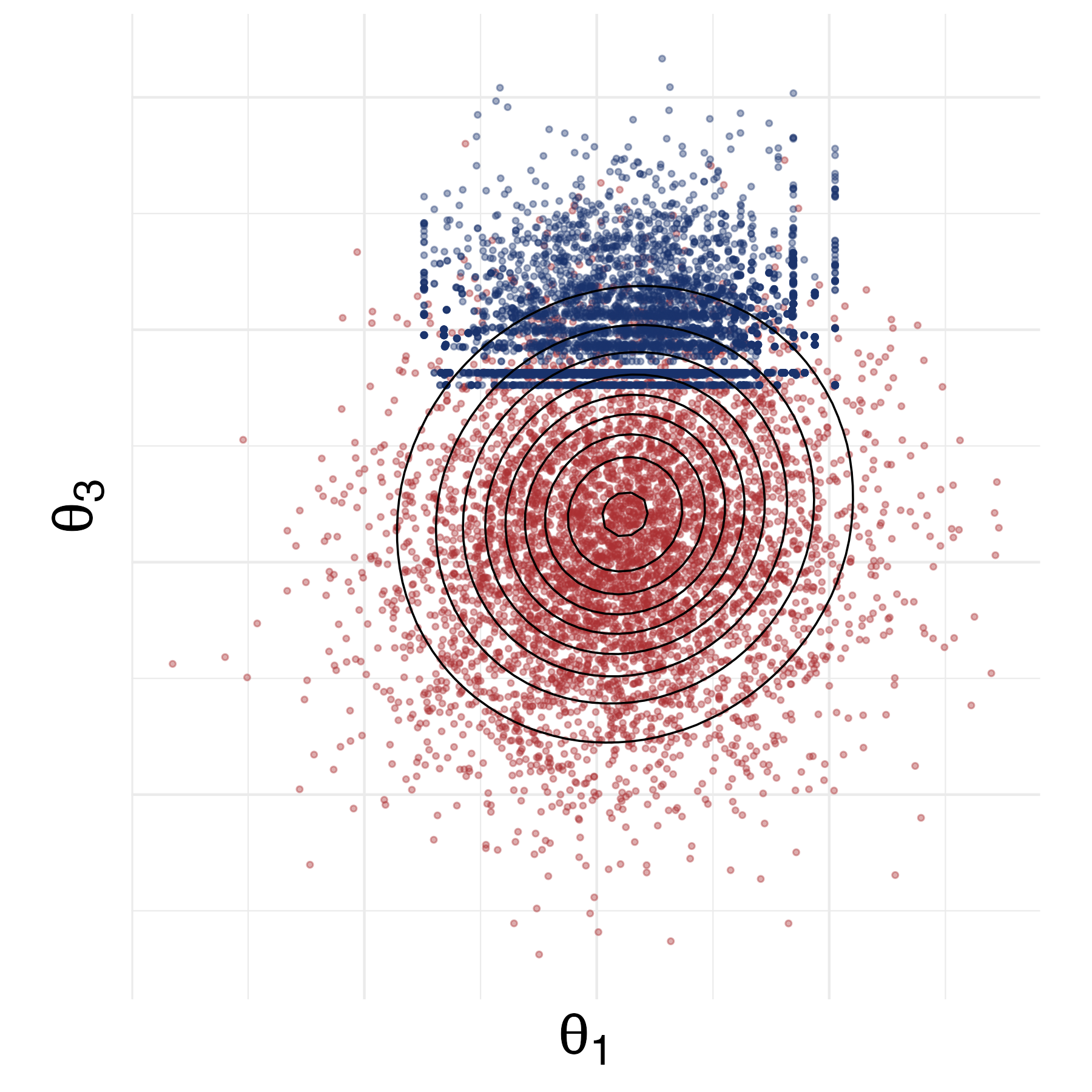}
			\caption{\centering Two-dimensional marginal $\theta_1, \theta_3 \mid \bm{Y}$}
            \label{fig:post_comp_b}
		\end{subfigure}
		\hfill
		\begin{subfigure}{0.29\textwidth}
			\centering
			\includegraphics[width=\linewidth]{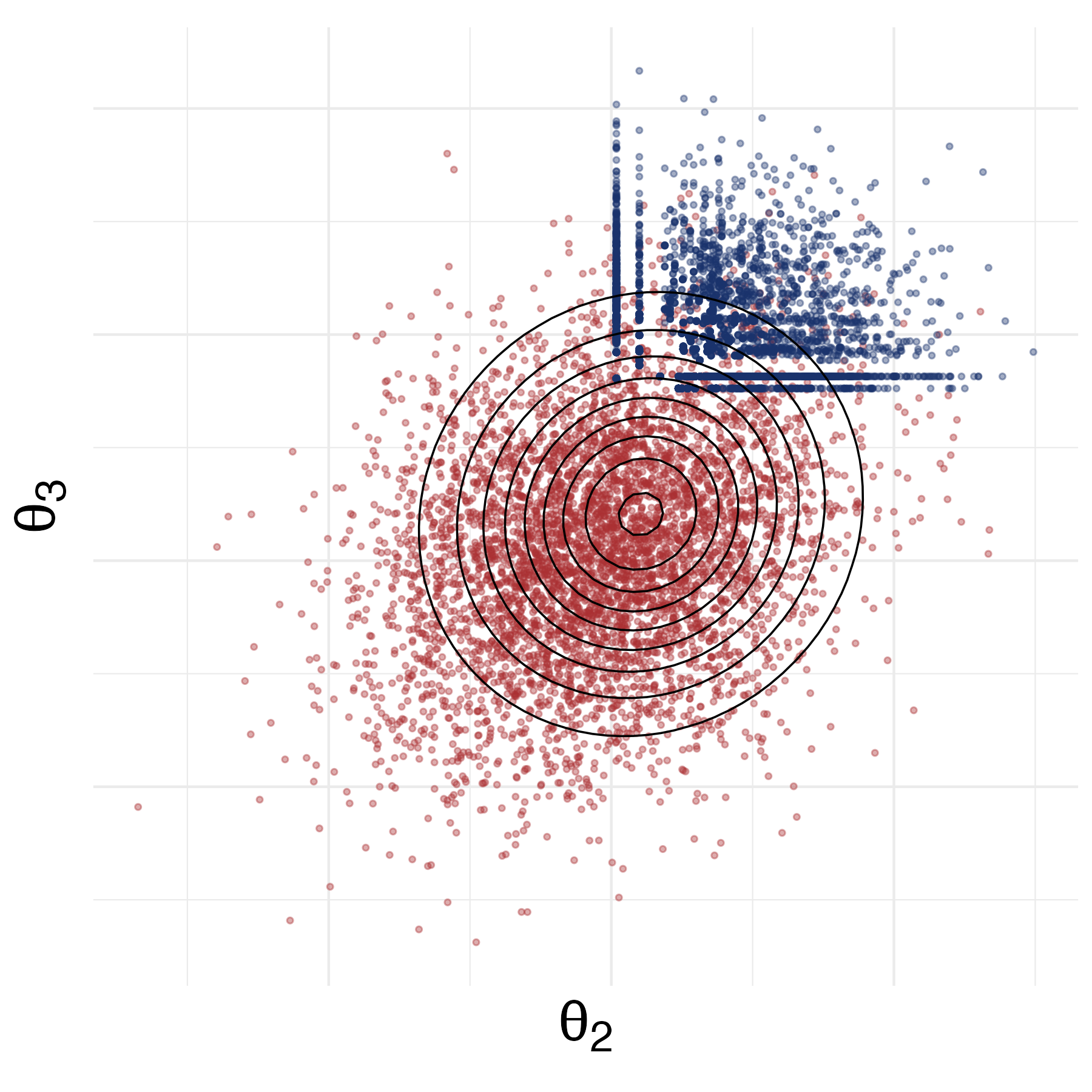}
			\caption{\centering Two-dimensional marginal $\theta_2, \theta_3 \mid \bm{Y}$}
            \label{fig:post_comp_c}
		\end{subfigure}
		
		\caption{Illustration of how the sampling prior displaces the software-generated sampler (a) and how the Two-Stage Fully Bayesian method fails to recover the posterior two-dimensional dependence structure and covariance (b,c). The estimates obtained by the Two-Stage Fully Bayesian method show strong dependence on the software-specific prior of the sampler, making the procedure sensitive to prior misspecification and less reliable when the prior shifts the generated draws away from the high-density regions of the target distribution.}
		\label{fig:hierarchical_output_bivariate}
	\end{figure}

    Figure~(\ref{fig:hierarchical_marginals})
    compares the true posterior marginals with the results obtained from both algorithms.
	The output of Stage 2 Normalizing Flows model closely matches the true $\theta_i \mid \bm{Y}$ and $\gamma \mid \bm{Y}$ distributions, capturing the shape and range of each marginal accurately.
	In contrast, the results for the Two-Stage Fully Bayesian method lack smoothness and appear abruptly truncated due to the strong dependence on
    the software prior. 
	As shown in Figure~(\ref{fig:post_comp_a}), if $P_S(\teta)$ pulls the software draws away from the posterior, the Two-Stage Fully Bayesian approach, using these samples as proposals, fails to cover the posterior regions which are not well represented by the sampling distribution. 
	This inaccuracy in the $\theta$ samples then propagates into the $\gamma$ estimates, since $\gamma$ is updated conditionally on the previously accepted $\theta$ values.
	Our approach avoids this issue because the specification of $h(\teta, \gamma)$ ensures that  $P_S(\teta)$ cancels out from the Stage 2 learning objective, leaving the reverse KL model unaffected by the software prior.
	Moreover, the proposed framework successfully recovers the two-dimensional dependence and covariance structure as shown in Figures~(\ref{fig:post_comp_b},\ref{fig:post_comp_c}).
	
\begin{figure}[h!]
		\centering
		\includegraphics[width=0.99\textwidth]{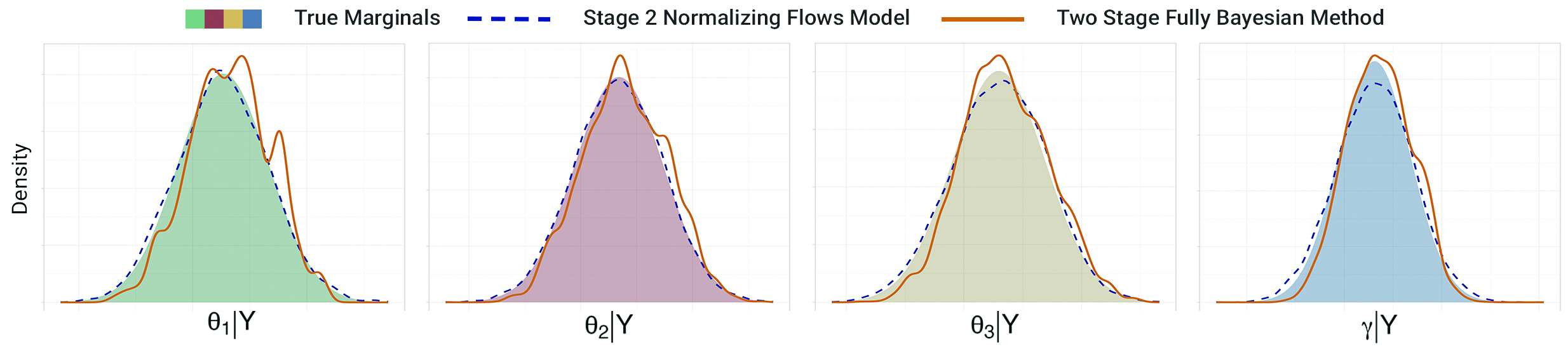}
		\caption{Selected posterior marginal profile comparison when the sampling prior is uniform in $\R^5$.}
		\label{fig:hierarchical_marginals2}
	\end{figure}

	We also considered a six-dimensional posterior ($J=6$) under a flat software prior, 
    $P_S(\teta) \sim 1$, to reduce the shift in the software draws. 
    The marginal density estimates presented in Figure~(\ref{fig:hierarchical_marginals2}) show 
    improvements in the Two Stage Fully Bayesian method. However, the Normalizing Flows output 
    provides more accurate marginal fits. Table (\ref{tab:nf_fb_error}) reports the estimation 
    errors in the marginal means and variances, indicating uniformly lower error for the proposed 
    framework. When examining pairwise relationships among the $\theta_i$ parameters in the 
    distribution of $\teta \mid \bm{Y}$. The Frobenius norm of the covariance matrix error is 0.016 for the Two-Stage Normalizing Flows method, compared to 1.846 for the Two-Stage Fully Bayesian approach.
	
	\begin{table}[h!]
		\centering
		\renewcommand{\arraystretch}{1.25}
		
		\begin{tabular}{c|c|c|c|c|c|c|c}
			\rowcolor[HTML]{eeebe7}
			\textbf{Model} & \textbf{Parameter} & $\teta_1$ & $\teta_2$ & $\teta_3$ & $\teta_4$ & $\teta_5$ & $\bm{\gamma}$ \\
			\midrule
			
			\rowcolor[HTML]{efece9}
			TSNF & Mean & -0.0265 & -0.0271 & 0.0051 & -0.0071 & -0.0015 & -0.0239 \\
			
			\rowcolor[HTML]{fbfaf8}
			TSFB & Mean & 0.2878 & 0.2556 & 0.2600 & 0.2405 & 0.2591 & 0.2523 \\
			
			\midrule
			
			\rowcolor[HTML]{efece9}
			TSNF & SD & 0.0162 & -0.0103 & 0.0144 & -0.0125 &  -0.0184  & -0.0133 \\
			
			\rowcolor[HTML]{fbfaf8}
			TSFB & SD & -0.2086 & -0.1067 & -0.1336 & -0.1089 & -0.1464 & -0.1793 \\
			
			\bottomrule
		\end{tabular}
		
		\caption{Comparison of estimation errors of marginal posterior parameters for the Two Stage Normalizing Flows method (TSNF) and the Two Stage Fully Bayesian (TSFB) Algorithm.}
		\label{tab:nf_fb_error}
	\end{table}
	
	Across these two hierarchical simulation settings
    with inaccessible likelihoods, the Two-Stage Normalizing Flows method consistently 
    outperforms the Two-Stage Fully Bayesian approach by providing more accurate and stable 
    estimates of the posterior shape, marginal distributions, and dependence structure, 
    regardless of the choice of the software prior. 
    In contrast, the Two-Stage Fully Bayesian algorithm is strongly affected by shifts in the 
    distribution of $P_S(\teta)$, which can induce bias and truncation in the resulting posterior 
    estimates. Even in the idealized setting of a flat software prior, 
    it fails to match the accuracy of Two-Stage Normalizing Flows framework in capturing 
    bivariate structures and key posterior features.
	
	% ----------------------------------------------

   % \section{Application of the Two Stage Normalizing Flows framework in Astrostatistics?}
    \section{Case Study: Hierarchical Inference in Astrophysics}
	\label{sec:realdata}

    In this section, we apply the proposed TSNF framework and the TSFB algorithm to a hierarchical inference problem from astrophysics. We are interested in the posterior distribution of the parameters for 30 white dwarf (WD) stars. Because the likelihood function is inaccessible, inference must be conducted using samples generated by a preexisting software suite known as BASE-9 \cite{BASE9}\footnote{We refer to the likelihood function as "inaccessible" because it is embedded within the BASE-9 software, and evaluating it involves evaluating computer models of stellar evolution that do not have closed-form expressions.
    %{\color{red} [add citations]}. Si et al. {\color{red} [make this a proper citation]}, when constructing the TSFB algorithm, refer to BASE-9 as a ``black-box'' software that cannot (practically) be modified to fit a hierarchical model. 
    BASE-9 does not allow for direct likelihood evaluations, and the allowable prior distribution formulations are limited in scope and do not admit hierarchical structure. Hence, the likelihood is inaccessible because we cannot directly evaluate it when fitting a hierarchical model, and must instead rely on the samples via BASE-9 from the individual posterior distributions of the 30 WD stars.}. The software uses an adaptive Metropolis MCMC procedure to sample the posterior distribution of the parameters of individual WD stars\footnote{BASE-9 was originally developed to explore the posterior distribution of star clusters---collections of stars that share certain physical properties in common. While we do not use that functionality in this work, the reader can find more information in the references cited above if interested. For this work, we focus solely on BASE-9's ability to sample from the posterior distribution of individual WD stars and do not explore its other uses.}; interested readers are directed to \cite{BASE9} for an overview of the software, \cite{vanDyk2009} and \cite{StenningPhD} for a description of the relevant astrophysics aimed at a statistical audience, and \cite{Stenning_multipop1} and \cite{Si_WDpaper} for an explanation of the MCMC routine used by BASE-9. 

    It is important to note that the true posterior distribution is unavailable, meaning that we cannot directly evaluate the joint posterior distribution of interest and must instead rely on the the two-stage approaches for drawing (approximate) samples. This makes direct accuracy comparisons of the two methods impossible, and so we do not try to demonstrate the superiority of the TSFN approach over the TSFB algorithm using quantitative criteria. Instead, the comparison between methods is necessarily qualitative, focusing on the structure, smoothness, and stability of the estimated posterior distributions. % for this realistic inference task.

    The setup for this inference problem is similar to the simulation study discussed in Section \ref{sec:hier-example} and is formulated based on the hierarchical model in Equation (\ref{hierarchical_model}). Here, $\theta_i$ denotes the base-10 logarithm of the age of the $i^{\text{th}}$ WD star, hereafter referred to simply as its {\it age}, which is modeled hierarchically through $\gamma$. The parameters $\bm{\xi}_i = (\xi_{i1}, \xi_{i2}, \xi_{i3}, \xi_{i4})$ 
    represent physical properties of the $i^{\text{th}}$ WD star and its observing conditions, specifically its {\it metallicity} ($\xi_{i1}$)---a proxy for the star's chemical composition, its {\it mass} ($\xi_{i2}$), the amount of its light blocked by interstellar matter, or {\it absorption} ($\xi_{i3}$), and, finally, the distance to the star as determined by its {\it parallax} ($\xi_{i4}$). 
    These parameters are inputs to a computer-based stellar evolution model, denoted by $G(\theta_i, \bm{\xi_i})$, which outputs the {\it photometric magnitudes} expected for a WD with parameters $\theta_i$ and $\bm{\xi_i}$. 
    Photometric magnitudes are a measurement of the brightness of the star in several wide wavelength bands, and, following \cite{vanDyk2009}, the observed photometric magnitudes are denoted by $\bm{Y}_{i}$, along with known measurement errors collected along the diagonal of the variance-covariance matrix $\Sigma_i$, where all off-diagonal terms are zero due to the assumed independence of the photometric magnitudes \cite{vanDyk2009}. With this, the heirarchical model is given by 
     %represents the vector of the remaining four physical parameters for $i = 1, \dots, 30$. The distribution $P_H(\bm{\xi_i \mid \theta_i})$ is replaced with $P_H(\bm{\xi_i})$ since absorption,
    %parallax, metallicity, and mass are assumed to be independent of log age values. The hierarchical model is described as
    \begin{equation}
		\begin{split}
			&
			\bm{Y}_{i} \sim N(G(\theta_i, \bm{\xi_i}), \Sigma_i),
			\quad i = 1,\dots, 30, \\
			&\bm{\xi_i} \sim P_H(\bm{\xi_i}),
			\quad i = 1,\dots, 30,\\
			&\theta_i \sim N(\gamma, \hat{\tau}^2),
			\quad i = 1,\dots, 30,\\
			&\gamma \sim N(9,0.5^2)
		\end{split} \label{real_hierarchical_model}
	\end{equation}
	%\new{where $ \hat{\tau} = 0.2783047$ is the standard deviation of the log-age values estimated from the software-generated samples across all 30 stars} 
    where $\hat{\tau}$ is fixed to be the standard deviation of the WD ages estimated from the software-generated samples across all 30 stars and $\gamma$ is the population-level parameter. Note that the distribution $P_H(\bm{\xi_i} \mid \theta_i)$ from Equation (\ref{hierarchical_model}) is replaced with $P_H(\bm{\xi_i})$ due to an assumed independence between $\theta_i$ and $\bm{\xi_i}$.
    
    In the above formulation, the function $G$ is not available in closed form, making the likelihood, $P_H(\bm{Y}_i\mid \theta_i)$, inaccessible and impossible to evaluate directly. Instead, from the BASE-9 software we have 15,000 draws of
  % 15000 samples of  $\teta_i\mid \bm{Y}_i$ for each of the 30 stars are available through the software sampler drawn from the distribution below:
    \begin{align}
		\nonumber
		&P_S(\theta_i, \bm{\xi}_i\mid \bm{Y}) \propto P_H(\bm{Y}_i\mid \theta_i, \bm{\xi}_i) P_S(\theta_i, \bm{\xi}_i),
		\quad \text{ with } \quad
		P_S(\theta_i, \bm{\xi}_i) = P_{S}(\theta_i)P_H(\bm{\xi}_i),
		\label{real_software_density}
	\end{align}
    for $i=1,\dots,30$, where $\bm{Y} = (\bm{Y}_1,\dots,\bm{Y}_{30})$ . Here, $P_{S}(\theta_i)$ denotes the software prior for $\theta_i$, taken to be uniform over a wide interval that corresponds to a stellar age of between 1 and 15 billion years. The priors on $\bm{\xi}_i$ are informed by existing astrophysical studies and reflect our assumptions about the plausible ranges of the parameters. The software prior for each $\xi_{i1}$ is a common normal prior, $N(-0.2,,0.5^2)$, used for all 30 stars. The base-10 logarithm of each $\xi_{i2}$ follows a normal prior $N(-1.02,,0.67729^2)$, which corresponds to the 
    %derived from the astrophysically motivated
    Miller and Scalo initial mass function \cite{miller_scalo_IMF}. For $\xi_{i3}$ and $\xi_{i4}$, star-specific normal priors are employed, with means and variances provided in Table \ref{tab:prior_abs_parallax} in Section \ref{sec:real_priors} of the Supplementary Materials.

\begin{figure}[h!]
		\centering
		\includegraphics[width=0.99\textwidth]{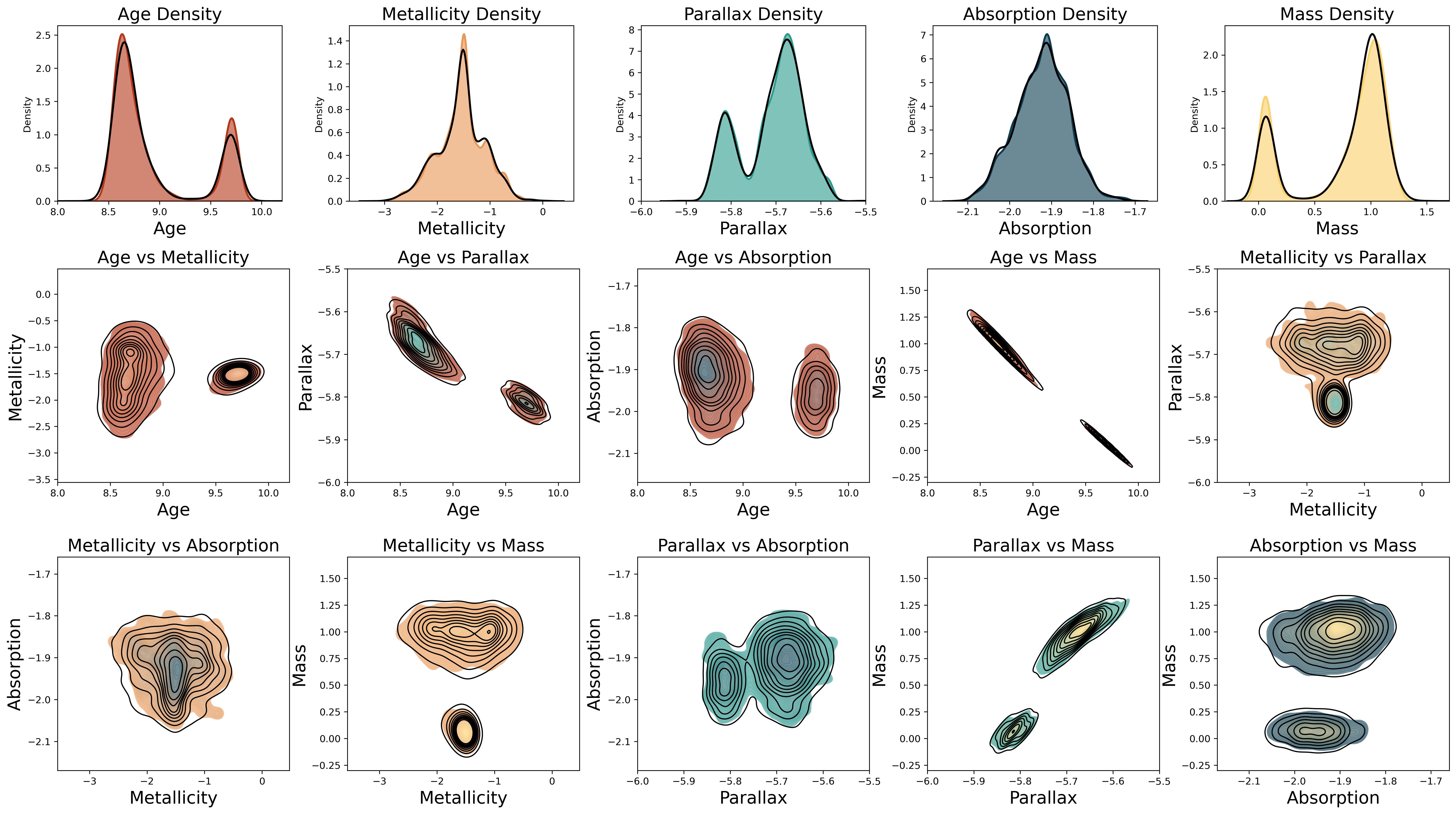}
		\caption{Marginal and bivariate density comparisons between the forward KL Normalizing Flows model and the software sampler $P_S(\theta_1 , \bm{\xi}_1\mid \bm{Y})$, where $\theta_1$ refers to the age parameter and $\bm{\xi}_1$ is the vector of metallicity, absorption, parallax, and mass. A natural logarithm transformation was applied to parallax, absorption, and mass in this analysis.}
		\label{fig:stage1_realdata_1}
	\end{figure}

    The goal of this analysis is to characterize the joint posterior distribution of all star-level parameters under the hierarchical model: 	
	$$P_H(\bm{\theta}, \bm{\xi}, \gamma \mid \bm{Y}) \propto P_H(\gamma)\prod_{i=1}^{30}P_H(\theta_i\mid \gamma)P_H(\bm{\xi})P_H(\bm{Y}_i\mid\teta_i,\bm{\xi}),$$
    where $\bm{\theta} = (\theta_1,\dots,\theta_{30})$ and $\bm{\xi} = (\bm{\xi}_1,\dots,\bm{\xi}_{30})$.
    To implement the TSNF method, we follow the procedure outlined in Section \ref{sec:NF_hierarchical}, and begin by fitting a forward KL Normalizing Flows model to approximate the density from the software sampler for each star separately. 
	Figure \ref{fig:stage1_realdata_1} displays the Stage 1 results for the first star in the dataset, illustrating how closely the Normalizing Flows model represents the distribution that generated the simulated draws. 
	The top row shows marginal densities for each parameter, with the solid black curves corresponding to the Normalizing Flows estimates and the colored regions indicating KDE of the software samples. The second and third rows display bivariate marginal densities, where black contour lines represent the fitted Normalizing Flows model and colored contours correspond to the empirical density of the software samples.
	In both cases, the output closely matches the structure of the software-generated samples, accurately capturing nonlinear dependence patterns, curvature, bimodality, range and spread within the data.
	
\begin{figure}[hb!]
		\centering
		
		% ---- Top figure ----
		\begin{subfigure}{\textwidth}
			\centering
			\includegraphics[width=1.02\textwidth]{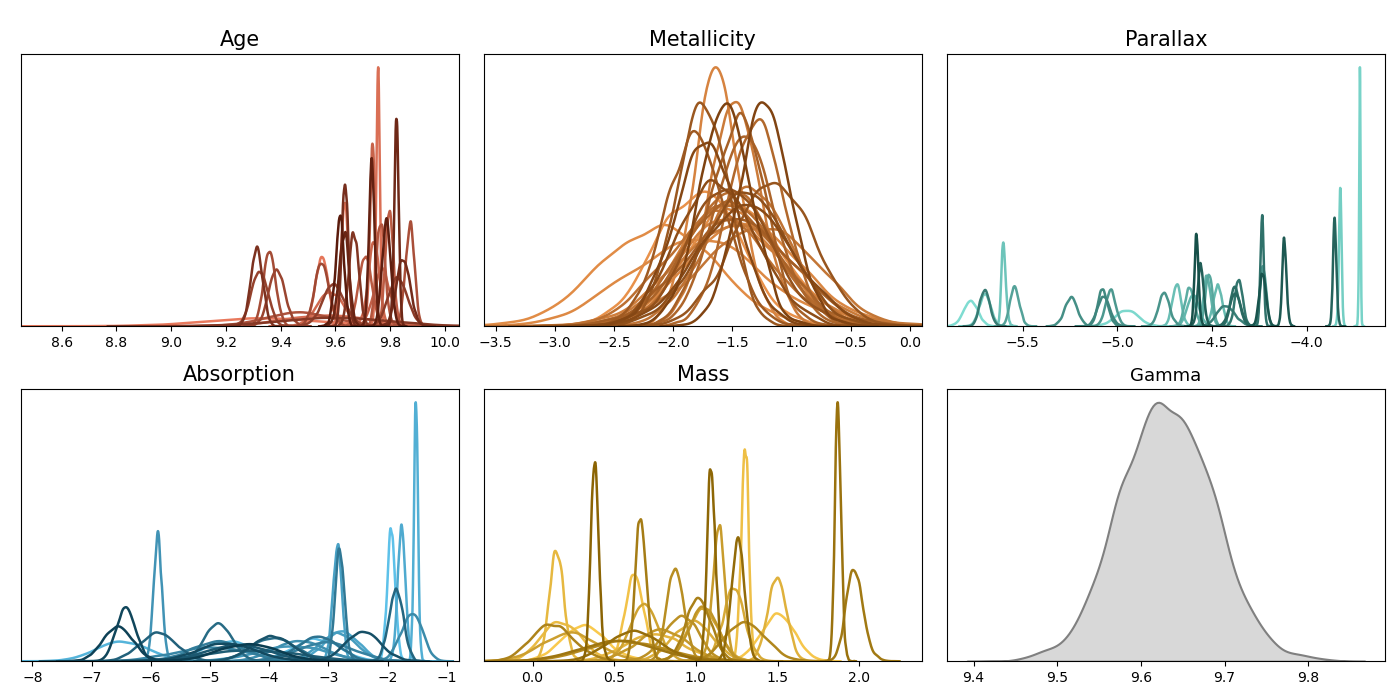}
			\caption{Two-Stage Normalizing Flows Method}
			\label{fig:stage2_tsnf}
		\end{subfigure}
		
		\vspace{0.5cm}
		
		% ---- Bottom figure ----
		\begin{subfigure}{\textwidth}
			\centering
            \includegraphics[width=1.02\textwidth]{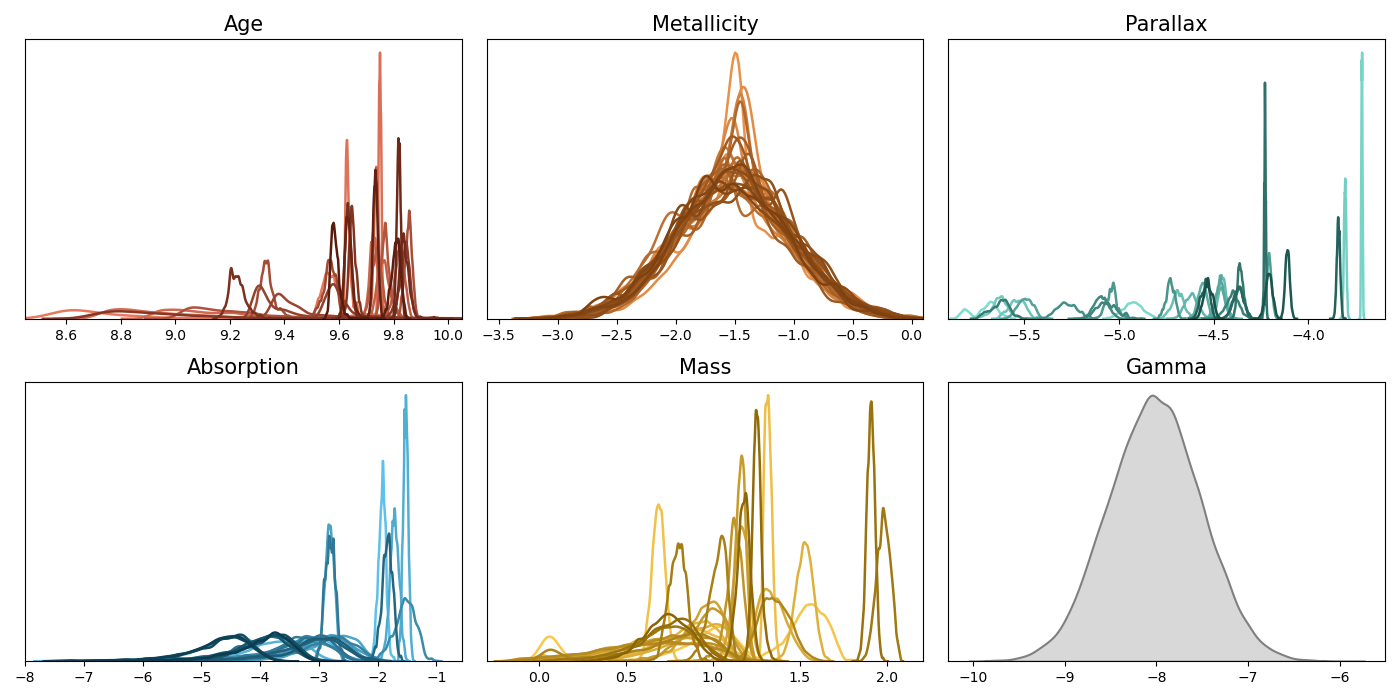}
			\caption{Two-Stage Fully Bayesian Method}
			\label{fig:stage2_tsfb}
		\end{subfigure}
		
		\caption{KDEs of the marginals of the hierarchical posterior obtained from the outputs of the Two-Stage Normalizing Flows (top) and Two-Stage Fully Bayesian (bottom) methods. Each subplot corresponds to one of the five star-level parameters, with individual curves representing marginal posterior KDEs for each of the 30 stars. The bottom-right subplot displays the estimated marginal posterior density of the population-level parameter $\gamma$.}
		\label{fig:stage2_realdata_comparison}
	   \end{figure}

	After obtaining the Stage 1 density estimates for all 30 stars, a reverse KL Normalizing Flows model is trained to approximate the full 151-dimensional joint posterior distribution. Marginal posterior estimates obtained from both the TSNF and the TSFB methods are shown in Figure \ref{fig:stage2_realdata_comparison}. Each subplot corresponds to one of the five star-level parameters, with individual curves representing marginal posterior estimates for each of the 30 stars. Overall, both methods produce similar posterior patterns.
	However, for parameters such as mass and absorption, the TSFB estimates exhibit substantial overlap across stars, whereas our TSNF method provides a more clear separation between star-specific marginals. This might potentially be an indication that the proposed approach has a superior performance given prior astrophysical knowledge, indicating meaningful variability among stellar parameters. Yet, our observations remain qualitative rather than definitive assessments of accuracy in such settings where the posterior distribution is difficult to evaluate directly, because it is not available in closed form, its density is computationally intractable or costly to evaluate, or it is generated by a black-box mechanism that does not allow evaluation at arbitrary points.
	
\begin{figure}[h!]
		\centering
		\includegraphics[width=0.99\textwidth]{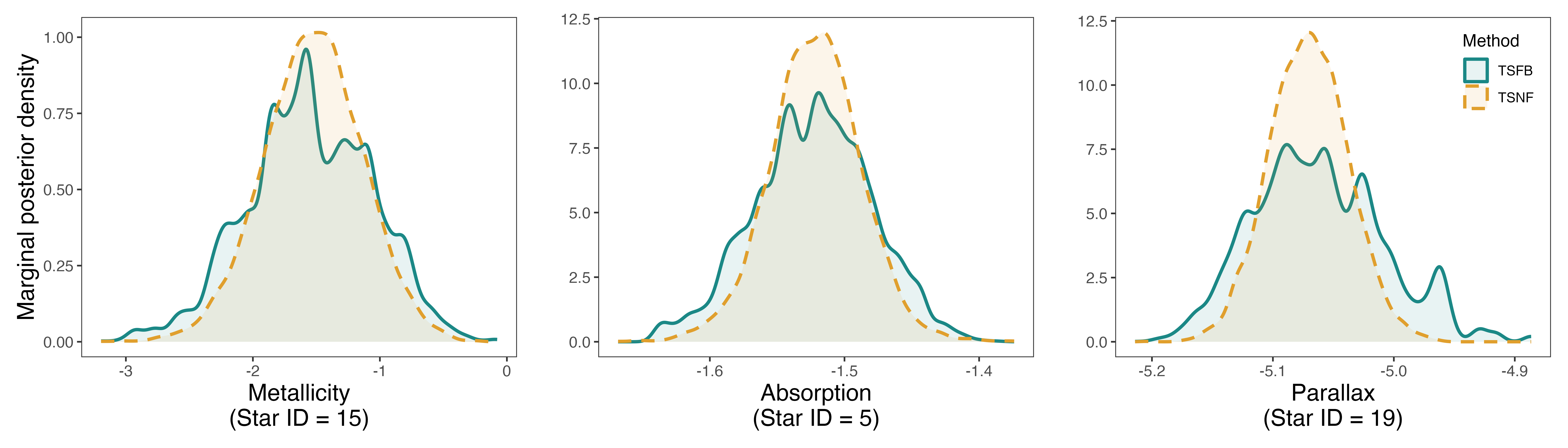}
		\caption{Selected posterior marginal profile comparison when the sampling prior is uniform in $\R^5$.}
		\label{fig:comparison}
	\end{figure}

    %Additionally,
    Another feature to highlight is that the marginal density estimates obtained via 
    %kernel density estimation 
    KDE from the samples produced by the TSFB method exhibit a noticeable lack of smoothness, particularly evident for the metallicity and parallax parameters in Figure \ref{fig:stage2_realdata_comparison}. This behavior is illustrated more clearly in Figure \ref{fig:comparison} for a selection of parameters for three different stars. A plausible explanation for this issue can be attributed to the fact that the TSFB method uses the limited software-generated samples as proposals within the Metropolis–Hastings step. As discussed in Section \ref{sec:hier-example}, repeated acceptance of the same values causes probability mass to accumulate at discrete locations in the parameter space. Since no proposals exist between these areas, the resulting KDEs exhibit sharp transitions and local irregularities in the posterior marginals.

	To conclude, this analysis illustrates the potential of the proposed framework in complex scientific applications involving real data.
	The TSNF method accounts for the effect of the software prior directly into the formulation of the target distribution in the second stage, preventing it from distorting the final posterior approximation. While the TSFB approach is faster and easier to implement, it may exhibit nonsmoothness, shifts, and bias in posterior estimates, particularly when the software prior is either misspecified or far from high-density regions of the target distribution. When the posterior distribution values are not available for validation, the proposed approach is expected to provide comparable estimates, with the potential to better explore the underlying target.

    % ----------------------------------------------
	\section{Conclusion}
	\label{sec:conclusion}
	
	This paper introduces a Two-Stage Normalizing Flows framework for probabilistic inference when piecewise information about the distribution of interest is available in different formats, such as samples from some components and structural assumptions and analytic expressions for others.
	The proposed approach reconstructs a tractable representation of the target by first learning sample-based components via a forward KL Normalizing Flow and then combining them with analytically known terms in a second reverse KL Normalizing Flows, supporting both efficient sampling and density evaluation. 
	This method relies on minimal assumptions and is broadly applicable in modern scientific applications where complete density evaluation or joint simulation is not feasible. 
	We demonstrated the approach in joint density inference with partially observed subvectors and in hierarchical models with inaccessible likelihoods, illustrating how the method can be used in applications where standard assumptions of classical inference procedures do not hold.
	
	The results presented in this paper show that the Two-Stage Normalizing Flows framework is sufficiently flexible to capture complex shapes, multimodality, and dependence structures, leading to improved parameter estimates and density approximations. 
	% \old{When implemented with stable flow architectures, the method scales favorably to higher-dimensional settings compared with most existing approaches. }
	While the primary focus of this paper was to introduce a general framework, future work can explore the properties of different Normalizing Flows layers tailored to specific problem structures. 
	Different transformation layers involve trade-offs between flexibility, numerical stability, and computational cost, and selecting flow architectures that require fewer layers and parameters may reduce training time and control instability in Jacobian evaluations. 
	Just as piecewise knowledge of the target distribution is used in constructing the model, such information could also guide the design of forward and reverse KL Normalizing Flows models to achieve more efficient and robust inference.
	
	Additionally, addressing the issue of model checking and goodness-of-fit assessment 
    remains an important consideration. For the models used in Stage 1, 
    samples from the target density at this stage are accessible and can be utilized to 
    evaluate the quality of fit through hypothesis testing, 
    empirical summary statistics, and visual diagnostics. However, neither 
    representative samples nor exact density values are available for the second stage, 
    making it difficult to formally validate the model. Designing practical diagnostic 
    tools with theoretical guarantees for such Normalizing Flows models could be a 
    promising direction for future research.
	
	Finally, the objective of this work was to introduce a general methodological framework 
    rather than to provide an exhaustive comparison with all alternative approaches. 
    Future work is required to
    investigate how the proposed Two-Stage Normalizing Flows algorithm compares with other 
    inference methods designed for settings with intractable likelihoods, such as 
    approximate Bayesian computation and other simulation-based inference methods. Such 
    comparisons would help to further clarify the strengths, limitations, and practical 
    domains of effectiveness of the proposed approach.
	% ----------------------------------------------

    \printbibliography
    
    \newpage
    \appendix
    \section{Supplementary Materials}
    \label{sec:appendix}
    Here we include additional details which were omitted from the main text.
    In Section~\ref{sec:TSFB_alg} we provide the complete algorithm
    for the Two-Stage Fully Bayesian approach. In Section~\ref{sec:app_sim}
    we derive the exact posterior distribution for our simulation study 
    in Section~\ref{sec:hier-example}, which
    allows quantitative comparison between our TSNF approach and the 
    existing TSFB algorithm. In Section~\ref{sec:real_priors}
    we provide the exact star specific software prior hyperparameters
    used.
    \subsection{Two-Stage Fully Bayesian Algorithm}\label{sec:TSFB_alg}
    In Algorithm~\ref{alg:two_stage_fb} we give the full algorithm for the TSFB approach. 
    This is taken from \cite{thesis}, and reproduced here for completeness.
    \begin{algorithm}
		\caption{Two-Stage Fully Bayesian algorithm.}
		\label{alg:two_stage_fb}
		\begin{algorithmic}[1]
			\State \textbf{Stage 1: } Generate software samples denoted by $(\teta, \bm{\xi})^{[S, t]}$ for $t = 1,\dots, T$. 
			\Statex
			\State \textbf{Stage 2:} Initialize by setting $(\teta, \bm{\xi})^{[TS, 0]} = (\teta, \bm{\xi})^{[TS, r]}$, where $r$ is a random number between $1$ and $T$.
			\For {$t = 1, \dots, T$: do}
			\State Draw $\psi^{[TS, t]}$ from $P_H(\psi)P_H(\teta^{[TS, t]}\mid \psi)$
			\State Given $\psi^{[TS, t]}$, update $(\teta, \bm{\xi})$ samples through a Metropolis Hastings algorithm where
			\begin{align*}
				&\text{Target: } P_H(\theta_i, \xi_i \mid y_i, \psi^{[TS,t]}),
				&\text{Proposal: } P_S(\theta_i,\xi_i \mid y_i).
			\end{align*}
			
			by recycling samples from Stage 1.
			\State Calculate acceptance probabilities $R_i$ via
			\begin{align*}
				R_i =&
				\frac{P_H((\theta_i, \xi_i)^{\text{Proposed}} \mid \psi^{[TS,t]}, y_i)/P_S((\theta_i,\xi_i)^{\text{Proposed}}\mid y_i)}
				{P_H((\theta_i, \xi_i)^{[TS,t-1]} \mid \psi^{[TS,t]}, y_i)/P_S((\theta_i,\xi_i)^{[TS,t-1]}\mid y_i)} \\
				& = \frac
				{P_H(\theta_i^{\text{Proposed}}\mid \psi^{[TS,t-1]})/P_S(\theta_i^{\text{Proposed}})}
				{P_H(\theta_i^{[TS,t-1]}\mid \psi^{[TS,t-1]})/P_S(\theta_i^{[TS,t-1]})}
			\end{align*}
			
			\State Set
			\begin{align*}
				(\theta_i,\xi_i)^{[TS,t]} = \begin{cases}
					(\theta_i, \xi_i)^{\text{Proposed}} & \text{with probability }\min{1,R_i}, \\
					\theta_i, \xi_i)^{[TS,t-1]} & \text{otherwise.}
				\end{cases}
			\end{align*}
			\EndFor
		\end{algorithmic}
	\end{algorithm}

	\subsection{Exact posterior for the simulation study}
    \label{sec:app_sim}
    The simulation study described in Section~\ref{sec:hier-example} provides us with a tractable
    posterior distribution, allowing quantitative comparison between our TSNF method and the 
    existing TSFB method.
	In particular, the marginal distribution of $\gamma \mid \boldsymbol{Y}$ is:
	$$P(\gamma \mid \boldsymbol{Y}) \propto \int P(\boldsymbol{\theta}, \gamma\mid \boldsymbol{Y})d\boldsymbol{\theta}, $$
    where for this problem we have that.
	\begin{align*}
		P(\theta_i\mid \gamma)P(y_i\mid \theta_i) \propto&
		\exp\left(-\frac{1}{2\tau^2}(\theta_i^2 -2\gamma\theta_i + \gamma^2)
		-\frac{1}{2\sigma^2}(\theta_i^2 -2\theta_i y_i)\right) \\
		=& \exp\left(-\frac{\sigma^2 + \tau^2}{2\sigma^2\tau^2}\left(\theta_i - \frac{\gamma\sigma^2 + y_i\tau^2}{\sigma^2+\tau^2}\right)^2\right)
		\exp\left(-\frac{1}{2}\left(\frac{\gamma^2}{\tau^2}-\frac{(\gamma\sigma^2 + y_i\tau^2)^2}{\sigma^2\tau^2(\sigma^2+\tau^2)}\right)\right) \\
		\Rightarrow \int P(\theta_i\mid \gamma)P(y_i\mid \theta_i) d\theta_i \propto & \exp\left(-\frac{1}{2}\left(\frac{\gamma^2}{\tau^2}-\frac{(\gamma\sigma^2 + y_i\tau^2)^2}{\sigma^2\tau^2(\sigma^2+\tau^2)}\right)\right)  \\
		\Rightarrow P(\gamma \mid \boldsymbol{Y}) \propto & \exp\left(-\frac{1}{2}\left(
		\gamma^2(\frac{J}{\tau^2}-\frac{J\sigma^2}{\tau^2(\sigma^2+\tau^2)})
		-2\gamma\frac{\sum y_i/J}{\sigma^2 + \tau^2}
		\right)\right). \\
		\propto& \exp\left(-\frac{J}{2(\sigma^2+\tau^2)}(\gamma^2 -2\gamma\bar{y})\right). \\
        \end{align*}
    Therefore, the posterior marginal of $\gamma|\boldsymbol{Y}$ is given by
    \begin{align*}
		 & \fcolorbox{gray}{appendix}{$\gamma \mid \boldsymbol{Y} \sim  N\left(\bar{y}, \frac{\sigma^2 + \tau^2}{J}\right).$}
	\end{align*}
	The distribution of $\gamma \mid \boldsymbol{\theta}, \boldsymbol{Y}$ would be
	\begin{align*}
		P(\gamma \mid \boldsymbol{\theta},\boldsymbol{Y}) \propto& \exp\left(-\frac{1}{2\tau^2}(J\gamma^2 + \sum \theta_i^2 -2\gamma \sum \theta_i)\right) \\
		=& \exp\left(-\frac{1}{2\tau^2/J}(\gamma - \bar{\theta})^2\right)\exp\left(-\frac{1}{2\tau^2}(\sum \theta_i^2 -J \bar{\theta}^2)\right).  \\
        \end{align*}
    Therefore, we have that
    \begin{align*}
		 & \fcolorbox{gray}{appendix}{$\gamma \mid \boldsymbol{\theta},\boldsymbol{Y}  \sim N\left(\bar{\theta}, \frac{\tau^2}{J}\right).$}
	\end{align*}
	
	Similarly, we can find the distribution of $P(\boldsymbol{\theta} \mid Y) = \int P(\boldsymbol{\theta}, \gamma\mid \boldsymbol{Y}) d\gamma$:
	
	\begin{align*}
		P(\boldsymbol{\theta}\mid \boldsymbol{Y}) \propto& \int P(\gamma)P(\boldsymbol{\theta}\mid \gamma)P(\boldsymbol{Y}\mid \boldsymbol{\theta}) d\gamma \\
		=& P(\boldsymbol{Y}\mid \boldsymbol{\theta})\exp\left(-\frac{1}{2\tau^2}\left(\sum \theta_i^2 -\frac{(\sum \theta_i)^2}{J}\right)\right)
		\int \exp\left(-\frac{J}{2\tau^2} (\gamma - \bar{\theta})^2\right) d\gamma \\
		\propto& \exp\left(
		-\frac{1}{2}\sum \theta_i^2(\frac{1}{\sigma^2}+ \frac{1}{\tau^2} - \frac{1}{J\tau^2}) 
		+\frac{\sum \theta_i y_i}{\sigma^2}
		+\frac{1}{J\tau^2}\underset{i\neq j}{\sum \sum}\theta_i\theta_j
		\right). \\
    \end{align*}
    This gives
    \begin{align*}
		 & \boldsymbol{\theta}\mid \boldsymbol{Y} \sim N(\boldsymbol{\mu}, \boldsymbol{\Sigma}),
	\end{align*}
	where
	$$\Sigma^{-1} = \begin{bmatrix}
		\frac{1}{\sigma^2} + \frac{1}{\tau^2} - \frac{1}{J\tau^2} & -\frac{1}{J\tau^2} & \cdots & -\frac{1}{J\tau^2} \\
		-\frac{1}{J\tau^2} & \frac{1}{\sigma^2} + \frac{1}{\tau^2} - \frac{1}{J\tau^2}& \cdots & -\frac{1}{J\tau^2} \\
		& \vdots&  \\
		-\frac{1}{J\tau^2} & \cdots & -\frac{1}{J\tau^2} & \frac{1}{\sigma^2} + \frac{1}{\tau^2} - \frac{1}{J\tau^2}
	\end{bmatrix} = \left(\frac{1}{\sigma^2} + \frac{1}{\tau^2}\right) I_J - \frac{1}{J\tau^2} 1_J1_J^T.$$
	Here, $I_J$ is the $J\times J$ identity matrix and every element of $J\times J$ matrix $1_J1_J^T$ is 1.
	Using the Sherman-Morrison formula to invert $\Sigma^{-1}$, we get:
	$$\Sigma = \left(\frac{\sigma^2\tau^2}{\tau^2 + \sigma^2}\right) I + \left(\frac{\sigma^4}{J(\sigma^2+ \tau^2)}\right)1_J1_J^T.$$
	% Also, 
	% $$\mu_i = \frac{\sigma^2 \bar{Y} + \tau^2 Y_i}{\sigma^2 + \tau^2}$$
	Therefore we have that
	$$\fcolorbox{gray}{appendix}{$\theta_i \mid Y_i \sim N\left(\dfrac{\sigma^2 \bar{Y} + \tau^2 Y_i}{\sigma^2 + \tau^2}, \dfrac{\sigma^2\tau^2}{\tau^2 + \sigma^2} + \dfrac{\sigma^4}{J(\sigma^2+ \tau^2)}\right)$,}$$
	$$\fcolorbox{gray}{appendix}{$\underset{i\neq j}{\mathrm{Cov}}(\theta_i, \theta_j) = \dfrac{\sigma^4}{J(\sigma^2+ \tau^2)}$.}$$
    
    % ----------------------------------------
    \newpage
    \subsection{Real Data Analysis: Software Priors}
    \label{sec:real_priors}
    \begin{table}[ht!]
	\centering
	\renewcommand{\arraystretch}{1.15}
	\begin{tabular}{c cc cc}
		\toprule
		\textbf{Star ID} & \multicolumn{2}{c}{\textbf{Absorption}} & \multicolumn{2}{c}{\textbf{Parallax}} \\
		\cmidrule(lr){2-3}\cmidrule(lr){4-5}
		& \textbf{Mean} & \textbf{SD} & \textbf{Mean} & \textbf{SD} \\
		\midrule
		1  & 0.150 & 0.010 & 3.458483  & 0.152516 \\
		2  & 0.040 & 0.020 & 7.436494  & 0.385762 \\
		3  & 0.010 & 0.005 & 24.298654 & 0.080329 \\
		4  & 0.030 & 0.015 & 22.214582 & 0.145184 \\
		5  & 0.220 & 0.010 & 3.601883  & 0.154057 \\
		6  & 0.180 & 0.015 & 3.936974  & 0.194410 \\
		7  & 0.060 & 0.030 & 9.363805  & 0.333163 \\
		8  & 0.060 & 0.005 & 11.669121 & 0.238129 \\
		9  & 0.060 & 0.030 & 10.021866 & 0.346149 \\
		10 & 0.050 & 0.025 & 11.520429 & 0.343922 \\
		11 & 0.050 & 0.025 & 10.489634 & 0.245862 \\
		12 & 0.050 & 0.025 & 11.502163 & 0.268556 \\
		13 & 0.220 & 0.040 & 4.136372  & 0.178349 \\
		14 & 0.040 & 0.020 & 14.900104 & 0.207197 \\
		15 & 0.060 & 0.030 & 6.536493  & 0.196926 \\
		16 & 0.040 & 0.020 & 8.890797  & 0.199346 \\
		17 & 0.060 & 0.005 & 5.169121  & 0.308389 \\
		18 & 0.040 & 0.020 & 6.188777  & 0.244796 \\
		19 & 0.020 & 0.010 & 6.623704  & 0.531326 \\
		20 & 0.020 & 0.010 & 12.159888 & 0.472029 \\
		21 & 0.160 & 0.015 & 3.665423  & 0.175912 \\
		22 & 0.050 & 0.025 & 12.877392 & 0.197370 \\
		23 & 0.050 & 0.025 & 14.572850 & 0.066508 \\
		24 & 0.020 & 0.010 & 12.740500 & 0.350400 \\
		25 & 0.020 & 0.010 & 12.724428 & 0.365909 \\
		26 & 0.020 & 0.010 & 21.480789 & 0.190042 \\
		27 & 0.010 & 0.005 & 10.702900 & 0.247406 \\
		28 & 0.010 & 0.005 & 16.394800 & 0.189728 \\
		29 & 0.010 & 0.005 & 14.869463 & 0.283223 \\
		30 & 0.020 & 0.010 & 10.824028 & 0.270755 \\
		\bottomrule
	\end{tabular}
    \caption{Star-specific software prior hyperparameters used for absorption and parallax.}
	\label{tab:prior_abs_parallax}
    \end{table}
    \newpage

	% \section{References}
    
\end{document}